\include{graphicsx}

\documentclass{emulateapj}

\slugcomment{accepted to ApJ September 5, 2006}

\shorttitle{Spitzer 24 Micron Observations of Open Cluster IC\,2391}
\shortauthors{Siegler et al.}

\begin{document}

\title{{\it Spitzer} 24 Micron Observations of Open Cluster IC\,2391 \\
and Debris Disk Evolution of FGK Stars}
\author{Nick Siegler\altaffilmark{1}, James Muzerolle\altaffilmark{1},
Erick T. Young\altaffilmark{1}, George H. Rieke\altaffilmark{1}, Eric E. Mamajek\altaffilmark{2}, David E. Trilling\altaffilmark{1}, Nadya Gorlova\altaffilmark{1}, \& Kate Y. L. Su\altaffilmark{1}}
\altaffiltext{1}{Steward Observatory, University of Arizona, 933 N. Cherry Ave., Tucson, AZ 85721 USA}
\altaffiltext{2}{Harvard-Smithsonian Center for Astrophysics, 60 Garden St., MS-42, Cambridge, MA 02138 USA}
\email{nsiegler@as.arizona.edu}
\begin{abstract}

We present 24\,$\micron$ {\it Spitzer}/MIPS photometric observations of the $\sim$\,50\,Myr open cluster IC\,2391. Thirty-four cluster members ranging in spectral type from B3-M5 were observed in the central square degree of the cluster. Excesses indicative of debris disks were discovered around 1 A star, 6 FGK stars, and possibly 1 M dwarf. For the cluster members observed to their photospheric limit, we find a debris disk frequency of 10$_{-3}^{+17}\%$ for B-A stars and 31$_{-9}^{+13}\%$ for FGK stars using a 15$\%$ relative excess threshold. Relative to a model of decaying excess frequency, the frequency of debris disks around A-type stars appears marginally low for the cluster's age while that of FGK stars appears consistent. Scenarios that may qualitatively explain this result are examined. We conclude that planetesimal activity in the terrestrial region of FGK stars is common in the first $\sim$\,50\,Myr and decays on timescales of $\sim$100\,Myr. Despite luminosity differences, debris disk evolution does not appear to depend strongly on stellar mass. 
\end{abstract}

\section{Introduction}

Nearly all stars are believed to form with primordial accretion disks,
but it is not clear whether the formation of planets is also a nearly-universal
process of stellar evolution. Answering this question would help us
understand the incidence of planetary systems in the Galaxy. Current
planet detection techniques, while continuously improving, all suffer
from some instrument sensitivity limitation. Many of the planets may
very well be insufficiently massive to be detected through gravitational recoil,
too faint against the glare of the central star to be imaged directly, too small to
significantly reduce the star's measured brightness, or positioned
unfavorably along the line of sight to produce a lensing event. 

Planetary debris disks, however, provide an additional approach. Debris
disks contain micron-sized dust grains predominantly
produced in collisions between larger-sized bodies (such as rocks). These dust grains are
heated by the parent star and re-radiate at longer wavelengths.
A key facet of debris disks is that the dust grains must be
short-lived compared to the age of the system given the efficiency of typical loss mechanisms like Poynting-Robertson drag
and radiation pressure (with timescales of 10$^{6}$ to 10$^{7}$ years). The dust must therefore be
regenerated either through a continuous collisional cascade or through
stochastic collisions. Therefore, the presence of dust implies the
existence of larger bodies that can collide and produce dusty debris \citep{bac93,lag00,zuc01}.
The largest of these bodies could be meter-sized up to planet-sized, and we may
well refer to them generally as planetesimals. {\em Therefore, any system with excess
thermal emission implies planet formation at least to the extent of forming planetesimals.} The ability to measure thermal emission in the mid-infrared is therefore a powerful technique in identifying systems in which planetary system formation has occurred or is occurring. However, since debris disks are cool, optically and
geometrically thin, and gas-poor, they are generally harder to detect
than the primordial, optically-thick accretion disks found around very
young stars ($\lesssim$\,10\,Myr). 

The {\em Spitzer} Space Telescope's unprecedented sensitivity in the
mid-infrared allows for the first time a statistical study of debris disks and their evolution across a wide spectral
range. Excesses detected at 24\,$\micron$ generally imply temperatures
on the order of 100\,K. This equilibrium temperature is achieved
in the vicinity of 1-5\,AU for spectral types FGK and 5-30\,AU for the more luminous B and A stars.
By probing these distances in the mid-infrared, we are therefore
studying potential planet-forming and planet-bearing regions
around other stars. Building on earlier work from the {\em Infrared Astronomy
Satellite} (IRAS) and {\em Infrared Space Observatory}
(ISO) which showed that the amount of dust in debris disks
steadily declines over time \citep[e.g.][]{spa01,hab01}, \cite{rie05}
showed
that more than half of A-type stars younger than $\sim$\,30\,Myr have mid-infrared
excess. This result implies that planetary system formation occurs frequently around stars a few times more massive than the Sun. However, the same result also shows that up to
$\sim$\,50$\%$ of the youngest stars have small or nonexistent excesses in the mid-infrared, pointing to a possible range of planetesimal formation and clearing timescales.

Can we expect similar behavior for lower-mass, longer-living,
solar-like stars? Both {\em IRAS} and {\em ISO} were in general not sufficiently sensitive to detect the photospheric
emission from lower-mass stars. Only with {\em Spitzer} have mid-infrared
surveys of lower-mass stars begun
\citep{gor04,you04,mey04,sta05,kim05,bei5b,che05,bry06,sil06,bei06,gor06,che06}. These studies conducted at 24 and/or 70\,$\micron$ have shown that debris disks exist around solar-like stars at a wide range of possible distances ($\sim$1-50\,AU) and temperatures ($\sim$10-650\,K) with an age-dependent frequency. It is one of the goals of this investigation to constrain this age-dependence. Continued surveys of stars with known ages at mid-infrared wavelengths will bring us nearer to understanding how debris disks evolve and ultimately will provide constraints on planet formation time scales.

In this investigation we use the 24\,$\micron$ channel on {\em
Spitzer} to study the incidence of debris disks in the open cluster IC\,2391. IC\,2391 is estimated to be 50$\pm$5\,Myr old \citep{byn04}, an age consistent with both theoretical \citep{cha01} and observational \citep{kle02} timescales of terrestrial planet formation. It is believed to be of intermediate size with $\sim$100-200 members. At a distance of 154\,pc \citep{for01}, the cluster is amongst the closest and best studied. Furthermore, its proximity allows for the detection of photospheric emission at mid-infrared wavelengths from low-mass stars. With little observed 24\,$\micron$ cirrus structure and visible extinction \citep[{\it E(B-V)}=0.006\,$\pm$\,0.005;][]{pat96}, IC\,2391 offers an attractive combination of age, distance, and background in which to study debris disks. 

The aim of this investigation is to measure the incidence of debris disks found around $\sim$\,50\,Myr stars across a broad range of spectral types. We discuss the ensemble properties of excesses
in IC\,2391 by placing our data in context
with other relevant samples. In
the process we begin characterizing the evolution of
debris disks around FGK stars, and compare this result
to that previously established for more massive
A stars.

\section{Observations and Data Reduction}

The Multiband Imaging Photometer for {\it Spitzer} \citep[MIPS;][]{rie04} was used to image a 0.97 square degree area 
(0.66$\degr$ $\times$ 1.47$\degr$) centered on IC\,2391 (RA 08:40:16.8, Dec -53:06:18.9; J2000) on 9 April 2004. The 24\,$\micron$ observations used the medium scan mode with 
half-array cross-scan offsets resulting in a total exposure time per pixel of 
80\,s.
The images were processed using the MIPS instrument team Data Analysis Tool \citep{grd05}, 
which calibrates the data, corrects distortions, and rejects cosmic rays 
during the coadding and mosaicking of individual frames. A column-dependent median subtraction routine was applied to remove any residual patterns from 
the individual images before combining them into the final 24\,$\micron$ mosaic.

While MIPS in scan-mode provides simultaneous data from detectors at 
24\,$\micron$, 70\,$\micron$, and 160\,$\micron$, this study is based on only 
the 24\,$\micron$ channel. The longer wavelength channels are insensitive to stellar 
photospheric emissions at the distance of IC\,2391 and in addition no 
cluster stars were detected at 70\,$\micron$ nor 160\,$\micron$.

We measured the 24\,$\micron$ flux density of individual sources in a 15$\arcsec$ aperture using the standard PSF-fitting photometry 
routine \texttt{allstar} in the \textsc{iraf} data reduction package \texttt{daophot}. We then applied an aperture correction of 1.73 to account for the flux density outside the aperture, as determined from the STinyTim 24\,$\micron$ PSF model (Engelbracht et al., in prep). Finally, 
fluxes were converted into magnitudes referenced to the Vega spectrum using a zero-point 
for the [24] magnitude of 7.3\,Jy (from the MIPS Data Handbook, v2.3). Typical 1$\sigma$ measurement uncertainties for the MIPS 24\,$\micron$ fluxes are 50\,$\mu$Jy plus $\sim$\,5\% uncertainty in the absolute calibration (Engelbracht et al., in prep). The two are independent of each other and dominated by the latter. 

The 24\,$\micron$ mosaic of the central region of IC\,2391 is displayed in Figure \ref{fig1}. It likely covers a bit less than half of the spatial extent of the entire cluster \citep{byn01}. There is relatively little
background cirrus or extended emission in the field of view. As explained in \S3.3, MIPS is sensitive to detecting the photospheres of mid-K dwarfs at the distance of IC\,2391.

\section{Results and Analysis}

\subsection{IC\,2391 Cluster Members Detected at 24\,$\micron$}

\begin{figure}
   \includegraphics[angle=0,width=\columnwidth]{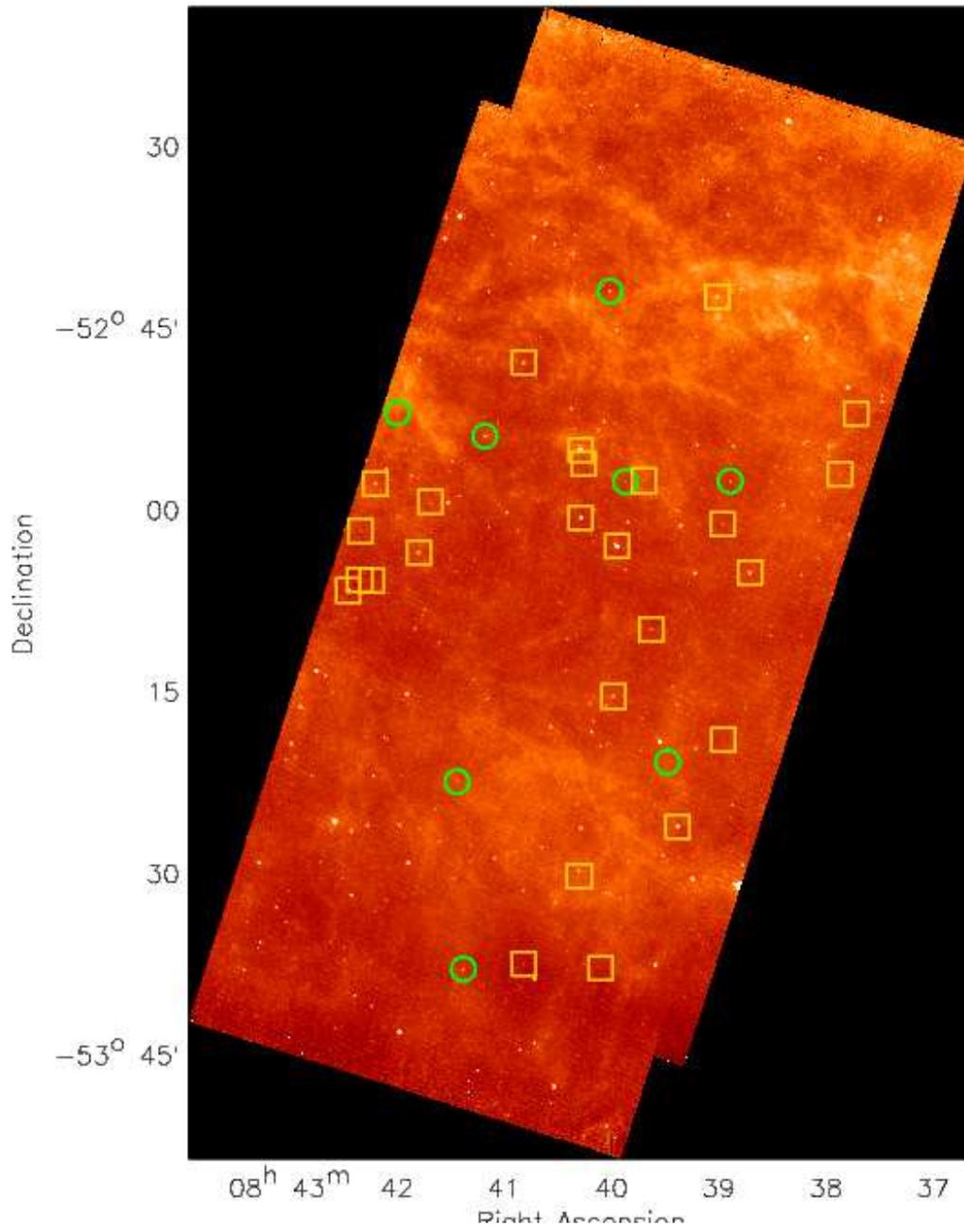}
\caption{ A 0.97 square degree mosaic of the central region of IC\,2391
taken with the MIPS 24\,$\micron$ channel. {\it Open circles}: debris disk candidates; {\it open squares}: cluster members with no apparent 24\,$\micron$ excess. The point-source FWHM is 5.7$\arcsec$ and the platescale is 1.25$\arcsec$/pixel. The image is displayed with a linear stretch and epoch J2000 celestial coordinates.
\label{fig1}}
\end{figure}

To determine the fraction of $\sim$\,50\,Myr stars possessing 24\,$\micron$ emission excess, 
we must match the detected sources in our mosaic to bona fide IC\,2391 cluster members. 
There are 1393 sources detected at 24\,$\micron$ in the {\em Spitzer}/MIPS 
mosaic (Figure \ref{fig1}) with a limiting magnitude of 11.7\,mag (0.15\,mJy). Using a two arcsec search radius, 505 of these sources matched objects in the 2MASS All-Sky Point Source Catalog \citep{cut03} providing both corresponding near-infrared photometry and standardized 2MASS celestial coordinates. It is expected that all IC\,2391 cluster members detected at 24\,$\micron$ in the MIPS mosaic will have corresponding 2MASS 
detections since the faintest known cluster members in the literature have $K_{s}$$\approx$\,14.5\,mag \cite[M5-M7;][]{byn04}, (2MASS $K_{s}$ 
sensitivity limit is $\simeq$\,15.3\,mag).

To obtain {\it V} band magnitudes and proper motions for cluster member selection, we ran the list of 505 sources through the United States Naval Observatory 
Flagstaff Station (USNOFS) image and catalog archive database NOMAD \cite[Naval Observatory Merged 
Astrometric Dataset\footnote{http://www.nofs.navy.mil/data/fchpix};][]{zac04}. This database selects for each 
source the "best" astrometric and photometric data chosen from its catalogs\footnote{for catalog 
details and references see http://www.nofs.navy.mil/nomad/nomad$\_$readme.html} and merges the 
results into a single dataset. Due to the cluster's distance, most of the sources had measured USNO-B1.0 \citep{mon03} or UCAC2 \citep{zac05} proper motions. In the cases where {\it V} magnitudes were not available through the database, we used alternate catalogs through the VizieR Search Service or photometry directly from literature sources listed hereafter.

To our list of 505 stars with {\it V, J, H, K$_s$}, and [24] photometry, we applied the following membership criteria in sequential order (numbers in parenthesis indicate the number of sources that still remained after the criterion was applied):
\begin{itemize}
\item object positions located on the stellar main sequence locus of a dereddened {\it J-H} vs {\it H-K$_s$} color-color diagram that indicate membership. (228)
\item object positions located on dereddened {\it V} vs {\it V-K$_s$} (Figure \ref{fig3}) and {\it K$_s$} vs 
{\it J-K$_s$} color-magnitude diagrams (CMDs) that indicate membership. (111)
\item proper motions within two sigma of the cluster mean \cite[$\approx$\,95\% of true members; estimated through a $\chi^{2}$ 
comparison to the mean Hipparcos cluster motion\footnote{\cite[$\mu_{\alpha}$cos$\delta$=-25.06$\pm$0.25\,mas/yr, $\mu_{\delta}$=22.7$\pm$0.22\,mas/yr;][]{rob99}} which includes the object's proper motion uncertainty 
and an assumed intrinsic velocity dispersion of 1\,mas/yr, where 1\,mas/yr $\approx$\,0.7\,km/s; p.71 in][]{bev92}. Using this criterion is expected to result in only $\approx$\,5\% of bona fide cluster members being rejected. (26)
\end{itemize}

The evolutionary models of \cite{bar98} and \cite{sie00} were used to determine the mean cluster CMD isochrones and color-color positions for 50\,Myr old stars placed at the distance of IC\,2391 (154\,pc). We selected candidate members using a band 1\,mag in apparent magnitude on either side of the mean theoretical isochrones and 0.1\,mag in color-color positions. The selection bands are sufficiently broad to take into account photometric, distance, age, binarity, and model uncertainties; reddening is not a factor here. However, with the cluster being close to the Galactic plane (b=-6.90), there is no clear separation between the field stars and the location of the cluster isochrones. We reduce the interloper contamination of our sample by using the combination of photometric and kinematic measurements as listed above. For later-type stars, however, the evolutionary models appear to diverge at {\it V-K${_s}$}$\gtrsim$\,4.4 and hence we used the spectroscopically-confirmed mid-M dwarfs from \cite{byn04} to define an empirical cluster sequence for later-type members.

Only those sources that satisfied all of the criteria were classified as members and are included in the statistics. From the original 505 sources, 26 met all three criteria for membership. As a consistency and completeness check, we compared our list to probable cluster members from the literature lying in the MIPS field of view in Figure \ref{fig1}. While there have been many cluster membership investigations of IC\,2391 measuring proper motions, 
optical and near-infrared photometry, radial velocities, rotational velocities, X-ray emission, spectral 
classification, and spectral youth diagnostics, there 
is no single complete listing. The Open Cluster Database\footnote{http://www.noao.edu/noao/staff/cprosser} 
as provided by Prosser \& Stauffer is composed of both members and candidate members 
extracted from the literature up until 1997. Later cluster membership references come from \cite{sim98}, \cite{pat99}, \cite{byn01}, \cite{for01}, \cite{ran01}, and \cite{byn04}. All 26 sources from our analysis were also classified in the literature as probable or possible cluster members with all but two (ID\,7 and 11) having spectral confirmation. Having satisfied our membership criteria, we are confident that these 26 sources are bona fide IC\,2391 cluster members and we list them in Tables \ref{tbl-1} and \ref{tbl-2}.

In addition, there were seven sources which were originally deselected due to proper motions slightly exceeding our $\chi^{2}$ criterion or not measured but are cited as probable cluster members in the literature. All seven - ID\,8, 10, 13, 21, 27, 29, and 32 - have photometry consistent with membership according to our first two criteria. ID\,8 (HD\,74009) is an F3 star whose proper motion we recalculated using additional catalogue points and now satisfies the third criterion. Using an 88-yr baseline, we also verified that ID\,8 is part of a 5.5$\arcsec$ binary whose companion, ID\,7, is also detected in our 24\,$\micron$ image. The companion independently satisfies the first two membership criteria but with large V band uncertainty. We thus add ID\,7 in addition to the original 26 sources. ID\,10 \cite[PP\,07;][]{pat99} is an M5 dwarf that we discuss in more detail in \S3.6. ID\,13 (SHJM\,6=VXR\,PSPC\,12), ID\,27 (SHJM\,8=VXR\,PSPC\,38a=VXR\,17), and ID\,29 (SHJM\,9=VXR\,41) \citep{sta89,pat96} all have evidence of youth through strong Li I detections \cite[$\lambda$6707\AA;][]{ran01}. ID\,32 \cite[SHJM\,10=VXR\,47;][]{sta89,pat96} is observed to be an M2e
dwarf with a radial velocity within 1$\sigma$ of the cluster mean
\citep{sta97}, Li abundance \citep{ran01}, and
colors consistent with an early-M dwarf. However,
its position on a CMD 
requires it to be a near-equal mass multiple system. Lastly, ID\,21 (VXR\,PSPC\,18a), has no measured proper motion but is X-ray active \citep{pat96}, has youth signatures \citep{ran01}, and a radial velocity consistent with cluster membership \citep{sta97}. Thus we add these eight sources to raise our final sample size to 34 objects and include them in Tables \ref{tbl-1} and \ref{tbl-2}.

Spectral types in Table \ref{tbl-1} were obtained from the references previously mentioned or from SIMBAD; rotational velocities were obtained from \cite{sta89,sta97}, and fractional 
X-ray luminosities (log[L$_x$/L$_{bol}$]) were obtained from \cite{pat96}. For completeness, we list in Appendix A three objects that are classified in the literature as cluster members but through this study are shown to be unassociated. None of these sources had 24\,$\micron$ excess.

\subsection{Determining 24\,$\micron$ Photospheric Colors }\label{bozomath}

Our goal is to measure the fraction of IC\,2391 cluster members possessing evidence of debris disks by 
measuring 24\,$\micron$ flux densities in
excess of their expected photospheric emission. We now establish a photospheric base-line emission using a {\it V-K${_s}$} vs {\it K$_s$}-[24] color-color diagram to help identify these excess sources across a broad range of spectral types. 

When the sources are all of similar distance and spectral type, a {\it K$_{s}$} vs {\it K$_{s}$}-[24] 
CMD can identify potential 
excesses empirically \citep[see Fig.\,3 in a Pleiades disk study by][]{sta05}. Even without knowing the exact 
photospheric color at 24\,$\micron$, \cite{sta05}, in consideration of their uncertainties and relative excesses, designate stars with {\it K$_s$}-[24]\,$\gtrsim\,0.1$ as candidate debris disk sources. 
However, for a broader range in photospheric temperatures (color), results from \cite{gau06} show that the {\it K$_s$}-[24] photospheric color gradually reddens with 
cooler effective temperatures until abruptly turning redward for spectral types later than M0. We also see evidence of this behavior in a larger disk survey of the Pleiades conducted by \cite{gor06}.

The {\it V-K${_s}$} color is a good proxy for spectral type. The two bands are sufficiently separated in wavelength space to trace temperatures/spectral types as well as to break degeneracies that beset near-infrared colors near the K-M spectral type transition. The {\it K$_s$}-[24] color is a very good
diagnostic for mid-infrared excess since for stars earlier than M dwarfs, it is only weakly dependent on stellar temperature 
with both bands in the Rayleigh-Jeans regime. Using the {\it K$_s$}-[24] as a mid-infrared excess diagnostic, however, requires the $K_s$ band to be photospheric. Near-infrared excess is a diagnostic 
for optically-thick primordial disks probing emission from active accretion at radii $\lesssim$\,0.1\,AU. We find no evidence for near-infrared excesses from {\it H-K$_s$} colors in our sample. Furthermore, for stars older than 10\,Myr, the inner-most regions of primordial disks have largely dissipated \citep{hai01}.  Thus we conclude cluster member 
emission at {\it K$_{s}$} is photospheric and that significant {\it K$_{s}$}-[24] deviations from photospheric values imply the presence of a circumstellar component. 

The lack of a large sample of 
cluster members with no apparent 24\,$\micron$ excess in IC\,2391 across a wide spectral type range makes establishing a pure empirical
photospheric locus potentially inaccurate. There are only 16 
apparently-non-excess stars with {\it V-K$_s$}$\lesssim$\,3.0. On the other 
hand, a mid-infrared investigation of the Pleiades by \cite{gor06} offers a large homogeneous 24\,$\micron$
stellar sample of similar metallicity and distance as IC 2391 with only slightly older age. \cite{gor06} have identified 
57 Pleiades members with good quality detections at {\it K$_s$} and 24\,$\micron$ and no evidence of mid-infrared excess. We plot these 57 stars with colors 0.05$\leq${\it V-K$_{s}$}$\leq$3.0 in Figure \ref{fig4} to illustrate the relative tightness of the 
distribution. The age difference between IC\,2391 and the 
Pleiades will have negligible effect on the intrinsic {\it K$_s$}-[24] color with both wavelengths on the Rayleigh-Jeans side of the emission spectrum for the range of stars in which we are interested. The effect on the {\it V-K$_{s}$} color is less than 0.1\,mag according to a comparison of \cite{sie00} tracks. This is not surprising as 
pre-main sequence stars at 50\,Myr are already quite close to the main sequence. As we use the {\it K$_s$}-[24] color as the primary diagnostic for mid-infrared excess, a 10\% variation in {\it V-K$_s$} or less will have very little effect on identifying excesses when using Figure \ref{fig4}. Hence, for the {\it K$_s$}-[24]$\sim$\,0 regime (0.05\,$\leq$\,{\it V-K$_s$}$\leq$\,3.0), we use the larger sample of Pleiades members 
with no apparent mid-infrared excess to construct an empirical photospheric locus of stars on a {\it V-K${_s}$} vs {\it K$_s$}-[24] color-color diagram. This photospheric locus is applicable for spectral types 
from late-B to mid-K stars.

 To establish the photospheric locus for M dwarfs, we rely on the field M-dwarf survey of \cite{gau06}. We plot 
their points ({\it small open circles}) with matching {\it V} band magnitudes in Figure \ref{fig5}. The photospheric colors indeed 
turn redward with increased slope for stars with {\it V-K$_{s}$}$\gtrsim$\,3.6. We compare this locus with the predicted {\it V-K$_s$} colors for 50\,Myr stars \citep{sie00} with the spectral type/{\it K$_s$}-[24] relation from \cite{gau06} by a ({\it dashed line}).

\subsection{Sources with Apparent 24\,$\micron$ Excess} \label{bozomath}
Since the precision of the 24\,$\micron$ photometry in the Pleiades dataset is very similar to ours, we adopt the Pleiades 3$\sigma$ relative excess threshold ($\sigma$=0.05\,mag) as the criterion for thermal excess in our IC\,2391 study. A cluster member whose 24\,$\micron$ flux density exceeds its predicted photospheric emission at this wavelength by at least 15\% is a debris disk candidate. We refer to the ratio of observed to predicted flux density as the 24\,$\micron$ excess ratio and will discuss its evolution for FGK stars in \S4.3. We apply in Figure \ref{fig5} this photospheric locus for A-K stars along with the empirical photospheric locus for M dwarfs to our IC\,2391 sample. We uniformly deredden the stars using {\it E(B-V)}=0.006 and the IR reddening laws described in \cite{cam02} (assuming A$_{[24]}$\,$\sim$\,0). 

In the {\it V-K$_{s}$}$\leq$3.0 regime in Figure \ref{fig5}, we identify seven debris disk candidates: ID\, 2, 4, 13, 16, 24, 25, and 26. Of the three detected cluster M dwarfs, we observed only 1 obvious 
excess - the M5 dwarf ID\,10 (discussed further in \S3.6). For the color
regime not fitted by our models (3.0$<${\it V-K$_{s}$}$<$3.6), we assume in Figure \ref{fig5} a simple diagonal fit connecting the upper and lower regimes. This is
consistent with the positions of late-type stars ID\,5, 21, and 27 and results in one more
candidate excess source, ID 29. 

The images of the nine stars initially identified as debris disk candidates were 
visually inspected at 24\,$\micron$ to ensure they match a point spread function and do not 
include potential contamination by heated cirrus or background 
sources. In addition the stars were analyzed in the higher-resolution, near-infared 2MASS images\footnote{http://irsa.ipac.caltech.edu/applications/2MASS/IM/} for elongation due to possible tight binaries. Only source ID\,2 \cite[VXR\,02a, G9;][]{pat96} showed elongation in the 24\,$\micron$ image, due to a faint source appearing 6.6$\arcsec$ away. \cite{pat96} classify this faint companion object as a K3V but its near-infrared colors and position on the cluster CMD are more consistent with a background K giant and hence it is possible that the 24\,$\micron$ 
excess may be due to dust from an evolving background giant and not the cluster member ID\,2. 
Therefore, we do not classify ID\,2 as a debris disk candidate. 

In total, we identify eight cluster members with evidence of debris disks - one A star, six FGK stars, and one M dwarf. We {\it circle} the eight in Figures \ref{fig1} and \ref{fig5} 
and indicate them as 24\,$micron$ excess objects with a ``Y'' in Table {tbl-2}. 

The candidate debris disks presented here are the first observed in IC\,2391. A report of possible 25$\micron$ {\it IRAS} 
excesses around several of the cluster A and early-F stars
\citep{bac91} is not confirmed. Six members (all B stars) have 12\,$\micron$ {\em IRAS}
detections, all of which are photospheric. None have
IRAS detections at wavelengths $\geq60\,\micron$. In addition, none of the members have been previously detected 
with ISO. We summarize the overall number and 
frequency of excess objects by
spectral type in Table \ref{tbl-3} and discuss their interpretation
in \S4. 

The excess frequency of a sample is 
defined as the ratio of the number of excess sources to the total
number of sources. We include in this ratio only those IC\,2391 cluster members whose photospheres are detectable at 24\,$\micron$. We define this minimum flux density sensitivity as the 
completeness limit of our sample, calculated at the turnover in the {\it K$_{s}$} brightness distribution 
of all the sources in our MIPS image with 24\,$\micron$ detection ({\it K$_{s}$}$<$\,9.9). This brightness corresponds to spectral type $\sim$\,K4 in IC\,2391. Detections 
fainter than the completeness limit may be biased toward excesses.

While we identify 34 cluster members in our mosaic, eight are removed from our statistical analysis. Four have {\it K$_{s}$} magnitudes fainter than our
photospheric completeness level (ID\,10, 21, 27, 29) and another two are binaries whose individual components are outside the completeness limit (ID\,5 and 32). The last two, ID\,20 and 34, are both B3IV stars. It is known that early B stars are sufficiently hot to emit free-free emission that can also contribute 24\,$\micron$ flux \citep{cho87}. Hence, using a 15\% threshold, we report an overall excess frequency for the cluster of 
0.23$_{-0.06}^{+0.10}$ \cite[6/26; excess frequency uncertainties are reported throughout 
this report as 1$\sigma$ binomial probability distributions; see Appendix in][]{bur03}.

While there were no detections of IC\,2391 cluster members with the MIPS 70\,$\micron$ channel, we calculate upper flux limits at the 24\,$\micron$ source positions
using aperture photometry with a 1.83 pixel radius aperture and 1.927 aperture correction (STinyTim 70\,$\micron$ PSF model; Gordon et al., in prep). The 70\,$\micron$ upper limits are listed in Table \ref{tbl-2}.

\subsection{Contamination} \label{bozomath}
With a relatively flat background and little cirrus 
emission, the most likely contaminant in IC\,2391
is confusion from random line-of-sight positional overlap with distant optically-faint but 
infrared-bright galaxies and AGN, showing no sign of elongation in the MIPS image. What is the 
probability of such an accurate chance alignment? For example, with $\sim$2000 extra-galactic sources per 
square degree at 0.5\,mJy \citep{pap04}, a flux less than our completeness limit but greater 
than our detection limit, the probability of a chance 
background source observed within 0.5$\arcsec$ of a cluster member is 0.4\% [$\pi$(0.5$\arcsec$$^{2}$/(0.97$\times$60$^{2}$))$\times$2000$\times$32]. Except for ID\,10 and ID\,29, the 
faintest excess sources, the remaining excess candidates are at least a factor of two brighter 
and the extragalactic contamination is significantly lower.

Additionally, we looked for positional offsets between our 24\,$\micron$ excess sources 
and the corresponding 2MASS positions that could potentially indicate fake excess emission from 
a superimposed object. The average offset between all our MIPS objects and 2MASS positions 
for members from Table \ref{tbl-1} is 0.6\arcsec. Except for ID\,29, which is the second 
faintest excess candidate in the sample (and not included in our frequency statistics), the 
sources fall within a 1\arcsec \, circle centered on the 0.6$\arcsec$ systematic offset on a 
$\Delta$RA vs $\Delta$Dec plane.

\subsection{Debris Disk Correlations with Other Stellar Properties}\label{bozomath}

Due to both its age and proximity, IC\,2391 has been the subject of several rotational velocity 
and X-ray studies \citep{sta89,pat93,pat96,sta97,sim98,mar05}. To assess correlations 
between cluster members with and without evidence for debris disks with other stellar parameters, 
 we matched the cluster members in Table \ref{tbl-1} with 
information regarding binarity, rotation ({\it v}\,sin{\it i}), and X-ray luminosity (Log[L$_x$/L$_{bol}$]) 
from the literature. As also observed by \cite{sta05} and \cite{gor06} in their 
investigations of the Pleiades, we find no clear correlations between 24\,$\micron$ excess sources 
and any of these stellar properties.

\subsection{An Interesting Possible Cluster Member: PP\,7} \label{bozomath}

ID\,10 \citep[PP\,7;][]{pat99} is a spectroscopically measured M5
dwarf \citep{byn04} with an observed 24\,$\micron$ flux density
approximately 1.6 times the predicted photospheric level. PP\,7 is
the faintest star in Table \ref{tbl-1} to have a 24\,$\micron$ detection {\it
and} have an excess; however, it is 3 times above the MIPS 24\,$\micron$ detection limit. 
Its positions in both near-infrared and optical CMDs (Figure \ref{fig3}), as well as in the near-infrared color-color diagram, 
are consistent with membership, but only as a nearly equal-mass binary system. The Na\,I doublet ($\lambda8200$\AA) equivalent width and H$\alpha$ emission line are consistent 
with a young, late M dwarf. Using five 
astrometric positions we have calculated its proper motion to be 
$\mu_{\alpha}$cos$\delta$=-45.0$\pm$10.1\,mas/yr, $\mu_{\delta}$=20.8$\pm$10.7\,mas/yr. 
Compared to IC\,2391's mean motion \citep{rob99}, PP\,7 gives a kinematic $\chi^{2}$ 
of 3.9 for two degrees of freedom (i.e. 14$\%$ of bona fide cluster members should have proper 
motion values more deviant). Hence, PP\,7 appears to be kinematically consistent with membership in IC\,2391. 

PP\,7 appears to possess, however, a Li abundance anomaly. \cite{byn04} report a weak Li measurement (S/N$\sim$\,3) in its 
spectrum despite the star being about 2\,mag brighter than the empirical Li depletion 
boundary for the cluster at {\it K$_{s}$}. The possibility, as mentioned above, that the star may be an equal-mass binary brings it 0.75\,mag closer to the cluster Li depletion 
boundary.
While there exists the possibility that PP\,7 is a very young, nearby star but unassociated with IC\,2391, the simpler 
hypothesis may be that it is a cluster member whose Li has simply not burned as fast as 
other members of similar mass. A radial velocity measurement would most likely confirm membership. 
If proven to be a member, it would be only the fourth M dwarf older than 10\,Myr known to have a mid-infrared or submillimeter excess \cite[AU\,Mic, GJ182, 2MASS J08093547-4913033;][]{son02,liu04,you04}.

\section{Disk Frequency of IC\,2391 and Implications for Debris Disk Evolution}

We find eight IC\,2391 cluster members with spectral types between A and M possessing 24\,$\micron$ excess consistent with debris disks. There are two interesting aspects to our results: 1.) a possible dearth of 24\,$\micron$ 
excess around A-type stars; and 2.) an abundance of 24\,$\micron$ excess around FGK stars. 
We now discuss these results,  put them in the context of stars in clusters of similar and 
different ages, and interpret the implications for debris disk evolution.

\subsection{Dearth of Debris Disks Around Early-Type Stars in IC\,2391?}\label{bozomath}

Because of their high temperatures and luminosities, A stars are very efficient at illuminating the dust 
in debris disks, yet they are not so hot, as are early-B stars, that they excite gaseous disks that might masquerade as debris dust. 
Consequently, debris systems around A-type stars (B5-A9) have been studied particularly thoroughly. 
The largest surveys of this nature are \cite{rie05} (266 stars) and \cite{ksu06} (160 stars) who used both {\em Spitzer} and {\em IRAS} observations to study debris disk frequencies and evolution. Their samples, composed of both cluster 
members and field objects, range in ages between 5 and 850\,Myr and all the observations at 24-25$\mu$m are
sensitive to photospheric levels. 

How does the debris disk frequency of A-type stars in IC\,2391 compare to the larger surveys? 
To create a single robust sample to which we could compare our results, we 
combined the two samples of Rieke et al. and Su et al., removed the {\em IRAS} sources (which 
have larger scatter than the {\em Spitzer} data), used a common relative excess threshold 
criterion ($\geq$\,15$\%$), and only considered sources with estimated ages $\geq$\,10\,Myr. 
Whenever there was source duplication we used the more recent Su et al. 24\,$\micron$ excess ratio 
results due to improved reduction procedures and fitting to theoretical photosphere models. The combined data set 
consists of 276 stars which we place in arbitrary age bins. For the 31-89\,Myr age bin, the excess 
frequency is 0.44$_{-0.10}^{+0.12}$ (8/18). Thus, nearly half of the stars between 31-89\,Myr 
have evidence for debris disks. For the same B5-A9 spectral type range in IC\,2391,
we find an excess frequency of 0.10$_{-0.03}^{+0.17}$ (1/10). If we use the binomial distribution where the probability of ``success'' is 0.44$_{-0.10}^{+0.12}$, 
the probability that these two results are drawn from the same distribution is about 3\%.

There are three possibilities to explain this result: 1.) it may be just a statistical deviation,
given the only moderate probability that the difference is significant; 2.) it may signal that
the simple smooth decay with age used to characterize debris disks as demonstrated by \cite{rie05} is an oversimplification; or 
3.) it might indicate that the cluster environment has influenced the debris disk evolution.
The first possibility cannot be ruled out without observations of additional clusters at similar ages. Nevertheless, the
latter two are worth exploring because it is of interest to see whether variations in the debris
disk frequencies in clusters might be possible and what their causes might be.

How does this frequency compare to clusters of other ages? In Table \ref{tbl-4} we list the excess 
frequencies from {\it Spitzer} 24\,$\micron$ surveys of A-type stars from open clusters 
(and an OB association) with sample sizes $\geq$ 10 and plot them in Figure \ref{fig7}. In each cluster, the same 
relative excess threshold of 15$\%$ above the predicted photospheric emission has been used. Age estimates and their uncertainties are obtained from references within those listed in 
Table \ref{tbl-4}. 

The seven open clusters and the association  
closely follow the larger combined field and cluster sample. This is not unexpected 
with only the Pleiades and IC\,2391 not already included in the larger combined sample. While the 
fraction of stars with debris disks for the other clusters matches the overall behavior of the
entire sample, IC\,2391's disk frequency appears disproportionately low. Since its behavior does not
seem to be reflected in the other cluster results, it does not represent an overall departure from the
smooth decline in activity. That is, there is no evidence in favor of our second hypothesis. 

We now consider the third possibility, that the cluster environment might be responsible. 
Unlike the marginally smaller fraction of measured excesses found amongst the A-type stars in IC\,2391, Figure \ref{fig5} shows that excesses around the FGK stars appear to be more common. Is there a physical scenario that can explain the behavior of {\em both} stellar types? 

Metallicity is not a likely issue in this case because IC\,2391 has comparable metallicity \cite[Fe/H=-0.03$\pm$0.07;][]{ran01} to other clusters with higher frequency of A-type excesses \cite[ie. the Pleiades and M47;][]{nis88,ran01}. In addition, a dependence
of the incidence of excesses on metallicity has not been seen in solar 
analogs \citep{gre06}.

A hypothesis invoking mass segregation whereby the most massive stars settle towards the cluster center where stellar densities, and hence disk interactions, are highest does manage to explain why less-massive stars would have higher disk 
frequencies. However, theoretical simulations of the effects of primordial disk interactions in clusters the size of IC\,2391 ($\sim$100-200 members) during the early
period of gas dissipation predict low interaction rates \citep{ada06}. In addition, this phenomenon is not observed in other open clusters which should already have experienced mass segregation (such as the Pleiades and the Hyades).

Interestingly, \cite{sag89}, conducting a kinematic survey of proper motion data for eight clusters 
with ages ranging from 8 to 300\,Myr, found IC\,2391 to
be the {\em only} cluster that showed mass dependence as a function of
intrinsic proper motion dispersions. The higher mass stars in IC\,2391
were measured to have lower velocity dispersions than the lower mass stars suggesting mass 
segregation. Their results, however, suffer from several observational uncertainties including 
low proper motion accuracies and incomplete IC\,2391 cluster membership. Nevertheless, their claim is intriguing.

Another possibility centers on the photoevaporation of primordial disks. This process may be 
greatly accelerated around very luminous O stars \citep[e.g.][]{hol01}. 
The formation of such a star in a cluster is subject to small number statistics \citep{elm04} and it is possible that some clusters would have subjected their
members' primordial disks to this effect, while others would not. Given the short lifetimes
of O stars, the direct traces of the star would have disappeared by the age of IC 2391. 
To account for the differences between the A-type and FGK stars in this cluster, however, requires 
that either mass segregation play a key role in timescales short enough to expose the A-type stars and their primordial disks to the UV radiation of O stars {\em before} significant planetesimal formation, or that photoevaporation is less efficient inwards (towards the star), at radii where G stars radiate at 24 micron ($\sim$\,5 AU). Theoretical models differ on whether photoevaporation
could behave in this manner \citep{joh98,ric00,mat03,thr05}. Although relaxation timescales in clusters of the size of IC\,2391 are short enough \cite[$\sim$\,1\,Myr; p.\,190,][]{bin87} to support this hypothesis, the high primordial disk frequency $\sim$\,40$\%$ in NGC\,2244 ($\sim$\,2\,Myr), despite its including many O stars (Balog et al., in prep), would argue against it.

In summary, although it would be interesting to search for some of the effects we have discussed in other clusters, none of them gives a solid explanation for the behavior of IC\,2391. For the present, we need to assume that the lack of A-type star debris systems may just be due to statistics.

\subsection{Abundance of Debris Disks Around Solar-Type Stars in IC\,2391}\label{bozomath}

Seven cluster members in IC\,2391 with spectral types F to M
show evidence of 24\,$\micron$ excess. Considering just the spectral types within the completeness 
limit ($<$\,K4) and a 24\,$\micron$ relative excess threshold of 15\,$\%$, the excess frequency of FGK stars 
is 0.31$_{-0.09}^{+0.13}$ (5/16). In fact, even around solar-type stars (F5-K7), the excess frequency is $\sim$\,0.31 (4/13). Debris disks around solar-type stars 
in IC\,2391 appear to be common.

Unlike the A-star surveys of \cite{rie05} and \cite{ksu06}, there is to date no comprehensive mid-infrared study in the literature of 
the fraction of solar-type stars with debris disks over a broad range 
of ages. We list the excess frequencies for known {\em Spitzer} 24\,$\micron$ surveys sensitive to 
photospheric emissions of F, G, and possibly K stars in open clusters (and an OB association) in 
Table \ref{tbl-6}. Depending on distance, or target sample, 
the spectral type corresponding to the 24\,$\micron$ completeness limit brightness varies in each 
cluster. We state any assumptions made in estimating the debris disk frequency of each 
cluster in Appendix B. 

The evolution of the excess frequency of FGK stars is shown in Figure \ref{fig8}. The IC\,2391 results fill an age gap in the previous studies between 30 and 100\,Myr. The relatively 
large 24\,$\micron$ excess frequency
observed around FGK stars in IC\,2391 ($\sim$\,31$\%$) appears consistent for its age with an 
evolutionary decay model. There are two important results here: 1) the IC\,2391 result implies 
that planetesimals around solar-type stars are {\em still} undergoing frequent collisions 
in terrestrial planet-forming regions at $\sim$\,50\,Myr and 2) the fraction of FGK stars with 24\,$\micron$ excess 
appears to decay similarly to the trend seen for A-type stars.

While the uncertainties in the excess 
frequencies of the youngest systems are considerable, Figure \ref{fig8} clearly illustrates that planetesimal 
activity (collisions) within the terrestrial planet zones of FGK stars ($\sim$\,1-5\,AU) is common during at 
least the first 50\,Myr. This is consistent with the epoch of
terrestrial planet formation in our own Solar System \citep{kle02,jac05}.
In fact, planetesimal systems around FGK stars continue being collisionally active even within the first 
few hundred million years. Several hundred million years later, however, mid-infrared excesses in the $\sim$\,1-5\,AU regions become rare. A survey of 69 nearby, solar-like field stars with median age $\sim$\,4\,Gyr 
found only two with 24\,$\micron$ excess meeting or exceeding the 15$\%$ relative excess threshold \citep{bry06}. 
Also, the evolutionary decays presented in Figure \ref{fig8} are only aggregate behaviors; even at ages 
$\gtrsim$\,500\,Myr large episodic excesses, while quite rare, do
appear \citep[e.g.][]{bei05}. The rarity of large impacts in mature systems is consistent with both 
the models of \cite{ken04} and the cratering record of the terrestrial planets \citep{str05,gom05}. We illustrate this further in \S4.3.

Combining with the results of \cite{gor06}, the debris disk frequency around FGK stars 
follows a similar excess decay behavior as the more massive A-type stars. Despite the overall similarity of behavior, one could also conclude from Figure \ref{fig8} that the FGK decay characteristic time scale appears shorter than that of the A-type stars. This is possibly, however, 
a luminosity effect rather than a mass-dependent 
effect (since the more luminous A-stars heat up larger annuli of dust to the levels detectable at 24\,$\mu$m). Until longer-wavelength observations can probe larger distances around FGK stars for evidence of cooler dust, we conclude from Figure \ref{fig8} that {\it debris disk evolution does not appear to be strongly dependent on stellar mass}. 

Infrared excesses may originate from cascading collisions among asteroid-sized objects. However, some systems 
may be in a quiescent phase and not currently exhibit infrared excesses. In others, a wave of planet formation may have already passed through the regions probed at 24\,$\micron$ ($\sim$\,1-5\,AU) and planetesimal collisions may be occurring undetected at larger distances \citep{ken03}. This implies that the debris disk incidence reported here actually gives a lower limit to the fraction of stars possessing planetestimals or undergoing planetesimal formation. Since a third
of the stars have significant planetesimal-collision-generated excess emission at $\sim$\,50\,Myr, and the incidence of debris disks possibly rises at earlier ages, {\em it is likely that planetesimals form around the majority of solar-like stars}. However, even a more intriguing conclusion can be drawn if planetesimal collisions must be driven by gravitational perturbations from planet-sized objects, in which case it may be true {\em that 
most primordial disks around solar-like stars evolve to form planetary systems}.

\subsection{The Evolution of 24\,$\micron$ Excesses Around FGK Stars}

Understanding how the 24\,$\micron$ excess evolves over time around FGK stars may provide insights to the collision history in the terrestrial planet region. Taking luminosity differences into account, we now explore the evolution of the excess ratio, which we defined in \S3.3. The 24\,$\micron$ excess ratios of FGK stars 
in our IC\,2391 sample range between 0.9 and 2.0 with a median of 1.1. In Figure \ref{fig12} we plot these results along with the excess ratios of FGK stars from the other clusters listed in Table \ref{tbl-6} as a function of time. We also add two known solar-type stars not members of clusters but with measured mid-infrared excesses - HIP\,8920 \cite[300\,Myr, G0V;][]{son05} 
and HD\,12039 \cite[30\,Myr, G3/5;][]{hin06}. 
The upper envelope of the excess ratios appears to decrease rapidly within the first 
$\sim$\,25\,Myr followed by a gentler decay with characteristic timescale of $\sim$\,100\,Myr. This early rapid decay may represent the final clearing of disks that are transitional between the primordial and debris stages. 
By several hundred million years, the mean 24\,$\micron$ flux is close to photospheric. 

Over-plotted onto Figure \ref{fig12} are an inverse time ({\em solid line}) and an inverse time-squared
 ({\em dashed line}) decay. While more data at younger ages would better define the fit, the inverse time 
decay appears to best match the data's upper envelope at ages $\gtrsim$\,20\,Myr. Larger inverse powers overestimate the number of large excesses observed at earlier times. An inverse time decay is qualitatively consistent with collisions being the dominant grain destruction mechanism \citep{dom03}. \cite{che06} show that for a main sequence F5 star with dust mass between 0.001-1\,M${_\earth}$, the collision 
lifetimes for average-sized grains is {\it always} shorter than the Poynting-Robertson and corpuscular wind drag lifetimes at radial 
distances $<$\,100\,AU. \cite{dom03} and \cite{wya05} 
conclude that all observed debris disks are in the collision-dominated regime. 

The evolution of the observed excess ratios shown in Figure \ref{fig12} may be best interpreted as the evolution of dust generation from planetesimal 
collisions around FGK stars. As planetesimals are gravitationally scattered out of planetary systems, grow 
into Moon-sized objects or larger, or are ground down and removed via Poynting-Robertson drag, their fewer numbers result in less frequent but occasionally powerful collisions producing copious amounts of dust. Observationally, this translates to a decreasing mean 24\,$\micron$ 
flux excess ratio with occasional large outliers as shown in Figure \ref{fig12}. Potential examples of stars with excess appearing as spikes indicating that such collisions have occured recently (in the past $\sim$\,million years) in their planetary
systems are 2M0735-1450 \cite[80\,Myr, F9;][]{gor04}, HIP\,8920 \cite[300\,Myr, G0V;][]{son05}, and HD\,69830 \citep[2\,Gyr, KO;][]{bei05}.

Based on the rarity of objects with evidence of recent collisions and the generally low incidence of 24\,$\micron$ excess  we draw a conclusion similar to that of \cite{rie05}: {\it large collisions occur after the initial period of terrestrial planet formation as 
episodic, stochastic events}. The general decay in the excess ratio may very well correspond to the decline of the collision frequency within the inner parts of a planetary system analogous to the asteroid belt of our own Solar System. The larger excess ratios observed at earlier periods may be very reminiscent of our understanding of events 
in the early Solar System, in which an early period ($\lesssim$\,100\,Myr) of frequent and catastrophic 
collisions (e.g., 
the birth of the Moon) was followed by a declining rate of planetesimal impacts followed by one last 
brief period of heavy bombardment $\sim$600-700 Myr \citep{str05,gom05}. 

The general behavior of the FGK stars in Figure \ref{fig12} is remarkably
similar to the corresponding figure for A-stars \citep{rie05}. Besides the inverse-time decay of the excess ratio and episodic outliers at ages older than about a hundred million years, Figure \ref{fig12} also illustrates another similarity between the two populations - the fraction of stars that have no or little 24\,$\micron$ excess at a given age. This phenomenon occurs even for the youngest FGK stars despite their overall higher probability of having mid-infrared excesses. Analogous to the previous conclusion from A-type star studies \citep{spa01,dec03,rie05}, this possibly points to a distribution of planet formation and clearing timescales even within young stellar clusters. Given the uniform behavior, the range of 24\,$\micron$ excess 
measured over time should eventually provide quantitative constraints for theoretical models of planetary system evolution.

The results from numerical simulations investigating the evolution of dust generation from planetesimal 
collisions around solar-type stars by \cite{ken05} show a qualitative similarity to the observed behavior in Figure \ref{fig12}. Kenyon \& Bromley (see their Fig.\,4) show both a steady decline of the 24\,$\micron$ excess ratio after $\sim$\,1\,Myr due to 
the depletion of colliding bodies and episodic large increases due to individual massive collisions. However, there are a number of observed behaviors where the simulations do not yet match. The characteristic timescales of the simulations appear shorter than what is observed. For example, at the age of the oldest subgroups in Scorpius-Centaurus ($\sim$\,17\,Myr), the simulations show 24\,$\micron$ flux densities only 2-3 times photospheric as compared to the much larger ratios observed by \cite{che05} and shown in Figure \ref{fig12}. This difference is independent of possible contamination by remnant primordial (or transition) disks in the Scorpius-Centaurus sample. In addition, there are no large excess ratios (spikes) greater than two after 50\,Myr in the simulation results, unlike those of 2M0735-1450 and HIP\,8920 shown in Figure \ref{fig12}. Lastly, at no time before a hundred million years in the simulation does the 24\,$\micron$ excess ratio reach unity. Any theory of debris disk evolution will have to account for those stars that show no 24\,$\micron$ excess (within {\it Spitzer's} detection limits) at ages less than 100\,Myr. This is an important observed phenomenon discussed earlier that occurs in stars across a broad range of spectral types and ages. Why some stars, in particular the youngest ($\lesssim$30\,Myr), have mid-infrared excesses and others do not is still without clear explanation.

Additional mid-infrared observations of intermediate-mass stars with known ages should help further constrain the time scales and behavior of the evolution of debris disks and, ultimately, of planetary system formation.

\section{Conclusions}\label{bozomath}

We have conducted a photometric survey for dusty debris disks in the
$\sim$\,50\,Myr open cluster IC\,2391 with the
MIPS 24\,$\micron$ channel on {\it Spitzer}. This wavelength probes regions $\sim$\,5 to 30\,AU around A-type stars and regions $\sim$\,1 to
5\,AU around FGK stars. Due to the cluster's proximity,
fluxes of stars as late as spectral type $\sim$\,K4 can be measured down
to the photospheric level. Of the 34 cluster members detected, only
10$_{-3}^{+17}\%$ (1/10) of the A-type stars had 24\,$\micron$ flux
densities $\geq$\,15$\%$ that of the photosphere. This is
lower than the 31$_{-9}^{+13}\%$ (5/16) frequency measured for
FGK stars in the cluster as well as marginally lower than A-type stars located in
other young
clusters. However, it is possible that this difference simply reflects random statistical variations.

In comparison, 31$_{-9}^{+13}\%$ of the FGK stars in IC\,2391 have excesses. From their behavior, we find the following:
\begin{enumerate}
\item A high level of planetesimal activity (collisions) is still occurring in
terrestrial planet regions ($\sim$\,1-5\,AU) at $\sim$\,50\,Myr. 
\item The fraction of FGK stars with 24\,$\micron$ excesses decreases significantly
on timescales of $\sim$100\,Myr. This decay over time corresponds to the observed decline of the frequency of collisions within the inner parts of these systems analogous to the asteroid belt of our own Solar System.
\item The decay and variation of 24\,$\micron$ excess ratios around FGK stars is
very similar to that measured around A-type stars. Despite an overall decaying excess ratio
evolution, there are large fractions of young stars with no excess at
the youngest ages and rare large excesses at older ages indicative of episodic and stochastic events.
\item Despite differences in luminosity and in the annuli probed at 24\,$\micron$ between A-type and
FGK stars, debris disk evolution does not appear to be strongly influenced by stellar mass (for this range of spectral type).

\end{enumerate}

\acknowledgements
N.S. would like to thank Christine Chen, 
Scott Kenyon, and Mike Meyer for helpful discussions and John Stansberry, Brian
Patten, John Stauffer, David Barrados y Navasques for data pertaining
to cluster membership. Support for this work was provided by NASA through Contract Number 1255094 issued by JPL/Caltech. This research has made use of the Simbad and Vizier databases operated at CDS in Strasbourg, France; the Two Micron All Sky Survey (2MASS) data services, a joint project of the University of Massachusetts and the Infrared Processing Center/California Institute of Technology, funded by NASA and the NSF; the U.S. Naval Observatory (USNO) Naval Observatory Merged Astrometric Dataset (NOMAD). \textsc{iraf} is distributed by the National Optical Astronomy Observatories, which is operated by the Association of Universities for Research in Astronomy, Inc., under contract to the NSF. EEM is supported by a Clay Postdoctoral Fellowship from the Smithsonian Astrophysical Observatory.



\clearpage
\begin{deluxetable}{rcccrrcccl}
\tabletypesize{\tiny}
\tablecaption{General Characteristics of IC\,2391 Cluster Members with 24\,$\micron$ Detections  \label{tbl-1}}
\tablewidth{0pt}
\tablehead{
\colhead{} &
\colhead{} &
\colhead{} &
\colhead{} &
\colhead{{\it V}} &
\colhead{{\it V-K{$_s$}}} &
\colhead {vsini } &
\colhead {} &
\colhead {} &
\colhead {} \\
\colhead{ID} &
\colhead{RA (2000)} &
\colhead{Dec (2000)} &
\colhead{SpT} &
\colhead{(mag)} &
\colhead{(mag)} &
\colhead {(km/sec)} &
\colhead {binarity?\tablenotemark{a}} &
\colhead {Log(L$_x$/L$_{bol}$)} &
\colhead {common names\tablenotemark{b}} 
}
\startdata

 1 &     8:37:47.0 &   -52:52:12.4 &      F5 &  9.65 &  1.14 &     - &      N &      - &  HD\,73777  \\
 2 &     8:37:55.6 &   -52:57:11.0 &      G9 & 11.50 &  2.22 &     - &     N? &  -3.70 &          VXR02a   \\
 3 &     8:38:44.8 &   -53:05:25.4 &      B8 &  6.44 & -0.17 &     - &      Y &  -5.99 & HD\,73952,VXR04   \\
 4 &     8:38:55.7 &   -52:57:51.7 &      G2 & 10.29 &  1.61 &    34 &      N &  -4.47 &  SHJM1,VXR05  \\
 5 &     8:38:58.8 &   -53:19:12.7 &      M3 & 13.77 &  3.92 &  $<7$ &    SB2 &  -3.38 &          VXR06a   \\
 6 &     8:38:59.9 &   -53:01:26.3 &      F5 &  9.66 &  1.39 &  21.0 &      N &  -3.98 &           VXR07   \\
 7 &     8:39:02.8 &   -52:42:38.4 &      K: & 11.28 &  2.48\tablenotemark{c} &     - &      Y &  - &       HD\,74009B  \\
 8 &     8:39:03.4 &   -52:42:39.7 &      F3 &  8.78 &  0.98 &     - &      Y &  -3.80\tablenotemark{d} & HD\,74009A,VXR08   \\
 9 &     8:39:23.9 &   -53:26:23.0 &      B5 &  5.44 & -0.36 &     - &      N &  - &       HD\,74071   \\

 10\tablenotemark{e} &    8:39:29.6  &        -53:21:04.4  &           M5  &      17.31  &       5.69  &          -  &          Y?  &           -  &                 PP07   \\
11 &   8:39:38.8 &   -53:10:07.2 &      G: &  9.95 &  1.63 &     - &      N &  -3.68 & VXR11,CD-52-2482   \\
12  &          8:39:43.0  &        -52:57:51.1  &           F2  &       9.10  &       0.97  &          -  &           N  & $<$-5.14  &        HD\,74117   \\
     13  &          8:39:53.0  &        -52:57:56.9  &           K0  &      11.86  &       2.07  &         16  &           N  &       -3.63  &          SHJM6,VXR12    \\
14 &     8:39:57.6 &   -53:03:17.0 &    B5IV &  5.17 & -0.36 &     - &     SB & $<$-7.28 &       HD\,74146   \\
15 &     8:39:59.4 &   -53:15:39.4 &     A1p &  7.21 &  0.05 &     - &     SB &  -4.90 & HD\,74169,VXR13   \\
     16  &         8:40:01.6  &        -52:42:12.6  &           A7  &       8.48  &       0.55  &          -  &           N  & $<$-5.75  &            HD\,74145    \\
17 &     8:40:06.2 &   -53:38:06.9 &      G0 & 10.41 &  1.51 &    47 &      N &  -3.59 & VXR14    \\
18 &     8:40:16.2 &   -52:56:29.2 &      G9 & 11.84 &  2.24 &  22.0 &      N &  -3.25 & VXR16a   \\
19 &     8:40:17.5 &   -53:00:55.4 &      B7 &  5.55 & -0.34 &     - &     SB & $<$-6.95 &       HD\,74196   \\
20\tablenotemark{f}  &     8:40:17.6 &   -52:55:19.0 &    B3IV &  3.59 & -0.55\tablenotemark{f}  &     - &      N &      - & $o$\,Velorum,HD\,74195   \\
21 &     8:40:18.3 &   -53:30:28.8 &      K4 & 13.54 &  3.30 &     8 &      N &  -3.28 &          VXR18a   \\
22 &     8:40:48.5 &   -52:48:07.1 &      A0 &  7.26 &  0.05 &     - &     SB &  -4.92 & HD\,74275,VXR21   \\
23 &     8:40:49.1 &   -53:37:45.4 &      G1 & 11.15 &  1.88 &     - &      N &  -3.27 &          VXR22a   \\
     24  &          8:41:10.0  &        -52:54:10.6  &           F6  &       9.85  &       1.22  &         43  &        SB1?  &       -4.48  &       HD\,74340,VXR30    \\
     25  &          8:41:22.8  &        -53:38:09.2  &           F3  &       9.54  &       1.00  &          -  &           N  &           -  &            HD\,74374    \\
     26  &          8:41:25.9  &        -53:22:41.6  &          K3e  &      12.63  &       2.94  &         90  &         SB?  &       -3.00  &         SHJM3,VXR35a    \\
27 &     8:41:39.7 &   -52:59:34.1 &    K7.5 & 13.38 &  3.42 &    18 &     Y? &  -3.32 &    SHJM8,VXR38a   \\
28 &     8:41:46.6 &   -53:03:44.9 &      A3 &  7.55 &  0.63 &     - &      N &      - &       HD\,74438   \\
     29  &          8:41:57.8  &        -52:52:14.0  &         K7.5  &      13.57  &       3.30  & $<$15  &           N  &       -3.21  &          SHJM9,VXR41    \\
30 &     8:42:10.0 &   -52:58:03.9 &      A1 &  7.37 &  0.03 &     - &      N & $<$-6.69 &       HD\,74516   \\
31 &     8:42:12.3 &   -53:06:03.8 &      F5 &  9.88 &  1.52 &   67: &      N &  -3.95 &           VXR44   \\
32 &     8:42:18.6 &   -53:01:56.9 &     M2e & 13.96 &  4.07 &    95 &      Y &  -3.41 &    SHJM10,VXR47   \\
33 &     8:42:19.0 &   -53:06:00.3 &     B9p &  5.48 & -0.33 &     - &     Y? &  -6.67 &       HD\,74535   \\
34 &     8:42:25.4 &   -53:06:50.2 &    B3IV &  4.82 & -0.44 &     - &     SB &  -7.79 & HD\,74560,VXR48   \\

\enddata
\tablenotetext{a}{SB: spectroscopic binary, SB1: single-line spectroscopic binary, SB2: double-line spectroscopic binary.}
\tablenotetext{b}{VXR: \cite{pat96}, SHJM: \cite{sta89}, PP: \cite{pat99}}
\tablenotetext{c}{($K_s$)$_{2MASS}$ derived from ($K_s$)$_{Denis}$ and (J-$K_s$)$_{Denis}$ using the \cite{car01} transformation relation.}
\tablenotetext{d}{X-ray measurements from \cite{pat96} include emission from both ID\,7 and 8}
\tablenotetext{e}{Li abundance potentially inconsistent with membership; see discussion in \S3.6}
\tablenotetext{f}{The reported 2MASS K$_{s}$ photometry is flagged due to saturation. We do not include this source in any of the figures using K$_{s}$ nor is it included in the excess frequency calculations due to its early spectral type.}
\tablecomments{Celestial coordinates are from 2MASS}.
\end{deluxetable}

\clearpage
\begin{deluxetable}{rrrrrccc}
\tabletypesize{\tiny}
\tablecaption{Infrared Properties of IC\,2391 Cluster Members with 24\,$\micron$ Detections  \label{tbl-2}}
\tablewidth{0pt}
\tablehead{
\colhead{} &
\colhead{{\it K{$_s$}}-[24]} &
\colhead{[24]} &
\colhead{$\sigma$\,([24])} &
\colhead{24\,$\micron$ flux} &
\colhead{excess ratio\tablenotemark{a}}&
\colhead{24\,$\micron$}&
\colhead{70\,$\micron$ flux\tablenotemark{b}}\\
\colhead{ID} &
\colhead{(mag)} &
\colhead{(mag)} &
\colhead{(mag)} &
\colhead{(mag)} &
\colhead{(mag)} &
\colhead{excess?} &
\colhead{(mJy)} 
}
\startdata

 1          &  -0.06 &  8.57 &  0.03 & 2.71 & 0.94 & N &    $<$63 \\
 2          &   0.37 &  8.91 &  0.04 & 1.98 & 1.34 & N &      $<$82 \\
 3          &  -0.11 &  6.72 &  0.03 & 14.90 & 0.93 & N &     $<$108 \\
      4                &       0.25  &      8.43  &      0.03  &     3.09  &     1.23  &  Y &   $<$88  \\
5          &   0.14 &  9.71 &  0.04 & 0.95 & 0.86 &   N &   $<$102 \\
 6          &   0.07 &  8.20 &  0.03 & 3.83 & 1.05 & N &     $<$100 \\
 7                           &  -0.03\tablenotemark{c} &     8.83 &  0.04 & 2.14 & 0.93 & N &   $<$71  \\
 8          &   0.05 &  7.75 &  0.03 & 5.81 & 1.04 & N &     $<$52  \\
9          &  -0.20 &  6.00 &  0.03 & 28.98 & 0.86 & N &      $<$73 \\
     10               &       1.10  &     10.52  &      0.06  &     0.45  &     1.63  & Y &  $<$52  \\
 11         &   0.07 &  8.25 &  0.03 & 3.67 & 1.04 & N &      $<$73 \\
    12                &       0.14  &      7.99  &      0.03  &     4.65  &   1.13  & N &  $<$52 \\
    13                &       0.22  &      9.58  &      0.04  &     1.08  &     1.18  & Y &  $<$148  \\
14          &  -0.18 &  5.71 &  0.03 & 37.92 & 0.88 & N &     $<$109 \\
15          &  -0.08 &  7.24 &  0.03 & 9.28 & 0.95 & N &      $<$94 \\
16                &       0.61  &      7.32  &      0.03  &     8.59  &     1.77  & Y &  $<$159  \\
17          &   0.07 &  8.83 &  0.03 & 2.14 & 1.04 & N &      $<$75 \\
18          &   0.06 &  9.54 &  0.04 & 1.12 & 1.01 & N &      $<$74 \\
19          &  -0.14 &  6.03 &  0.03 & 28.27 & 0.91 & N &      $<$54 \\
20\tablenotemark{d}                           &  -0.08\tablenotemark{d} &  4.22 &  0.03 & 149.05 &   0.97 & N &  $<$136 \\
21          &   0.12 & 10.12 &  0.05 & 0.65 & 0.99 & N &        $-$ \\
22          &  -0.10 &  7.31 &  0.03 & 8.70 & 0.93 & N &     $<$102 \\
23          &   0.03 &  9.25 &  0.03 & 1.46 & 0.99 & N &        $-$ \\
    24                &       0.43  &      8.20  &      0.03  &     3.84  &     1.47  & Y &  $<$124  \\
    25                &       0.75  &      7.79  &      0.03  &     5.58  &     1.98  & Y &       -  \\
    26                &       0.26  &      9.44  &      0.04  &     1.23  &     1.18  & Y &   $<$115 \\
27          &   0.22 &  9.74 &  0.04 & 0.93 & 1.07 &  N &     $<$93 \\
28          &  -0.05 &  6.97 &  0.03 & 11.84 & 0.96 & N &      $<$67 \\
29                &       0.35  &      9.92  &      0.05  &     0.79  &     1.22  & Y &       -  \\
30          &  -0.11 &  7.44 &  0.03 & 7.70 & 0.93 & N &          - \\
31          &   0.03 &  8.34 &  0.03 & 3.37 & 1.00 & N &      $<$68 \\
32          &   0.30 &  9.59 &  0.04 & 1.07 & 0.96 & N &          - \\
33          &  -0.10 &  5.91 &  0.03 & 31.57 & 0.94 & N &          - \\
34         &  -0.20 &  5.47 &  0.03 & 47.48 & 0.87 & N &          - \\

\enddata
\tablenotetext{a} {Ratio of observed 24\,$\micron$ flux density to the predicted photospheric flux density at  24\,$\micron$.}
\tablenotetext{b}{Upper limits; the 24\,$\micron$ and 70\,$\micron$ mosaicks cover slightly different areas of the sky and hence some 24\,$\micron$ detections do not have 70\,$\micron$ upper limit measurements. }
\tablenotetext{c}{($K_s$)$_{2MASS}$ derived from ($K_s$)$_{Denis}$ and (J-$K_s$)$_{Denis}$ using the \cite{car01} transformation relation.}
\tablenotetext{d}{The reported 2MASS K$_{s}$ photometry is flagged due to saturation. We do not include this source in any of the figures using K$_{s}$.}
\end{deluxetable}

\clearpage
\begin{deluxetable}{ccccc}
\tabletypesize{\scriptsize}
\tablecaption{Fraction of IC\,2391 Cluster Members with 24\,$\micron$ Excess\label{tbl-3}}
\tablewidth{0pt}
\tablehead{
\colhead{} &
\colhead{$\#$ of} &
\colhead{$\#$ of}&
\colhead{} &\\
\colhead{} &
\colhead{Members in} &
\colhead{Excess} &
\colhead{Excess} &\\
\colhead{Spectral Type} &
\colhead{this Sample} &
\colhead{Stars\tablenotemark{a}} &
\colhead{Frequency\tablenotemark{b}} &
}
\startdata
B & 7 & 0 & $\sim$\,0.0 (0/5)\tablenotemark{c} \\
A & 5 & 1& 0.20$_{-0.08}^{+0.25}$ (1/5)\\
FGK & 19 & 6 & 0.31$_{-0.09}^{+0.13}$ (5/16)\\
M\tablenotemark{d} & 3 & 1 & - & \\
\hline\hline
B5-A9 (A-type) & 10 & 1 &0.10$_{-0.03}^{+0.17}$ (1/10)\\
F5-K4 (solar-type) & 13 & 4  & 0.31$_{-0.10}^{+0.14}$ (4/13)  
\enddata
\tablenotetext{a}{Excesses defined as sources with {\it K$_s$}-[24] colors lying on or redward of the 3$\sigma$ Pleiades photospheric baseline shown in Figure \ref{fig5}. 24\,$\micron$ excess fluxes are $\geq15\%$ above the mean photospheric flux as a function of spectral type.}
\tablenotetext{b}{Ratio of the number of excess stars to the total number of stars in a spectral bin. All sources included in the excess frequency have 24\,$\micron$ sensitivity to the photospheric flux (spectral type $<$\,K4).}
\tablenotetext{c}{Spectral types $<$\,B5 are removed from the excess frequency statistics due to possible free-free emission contamination.}
\tablenotetext{d}{All 3 systems are most likely binaries with integrated brightness sufficiently large to be detected at 24\,$\micron$; one, however, has a very strong 24\,$\micron$ excess. None of the 24\,$\micron$ detections are sensitive to photospheric emission.}

\end{deluxetable}

\clearpage
\begin{deluxetable}{ccccc}
\tabletypesize{\scriptsize}
\tablecaption{24\,$\micron$ Excess Frequencies of A-Type Stars from {\it
Spitzer}/MIPS Surveys\tablenotemark{a}\label{tbl-4}}
\tablewidth{0pt}
\tablehead{
\colhead{} &
\colhead{Excess Frequency} &
\colhead{Age (Myr)} &
\colhead{Sample Size} &
\colhead{Excess Frequency References}
}
\startdata
Up Cen Lupus & 0.44$_{-0.11}^{+0.12}$ & 16$\pm2$ & 16 & \cite{ksu06}\\
NGC\,2547 & 0.44$_{-0.10}^{+0.12}$ & 30$\pm5$ & 18 & \cite{you04}, Gorlova et al. (in prep)\\
{\bf IC\,2391} & {\bf 0.10$_{-0.03}^{+0.17}$} & {\bf 50$\pm5$} & {\bf
10} & {\bf This paper} \\
M47 & 0.32$_{-0.12}^{+0.10} $ & 80$\pm20$ & 31 & \cite{rie05}\\
Pleiades & 0.25$_{-0.07}^{+0.12}$ & 115$\pm20$ & 20 & \cite{gor06}\\
NGC\,2516 & 0.25$_{-0.05}^{+0.07}$ & 150$\pm20$ & 51 & \cite{rie05}  \\
Hyades & 0.09$_{-0.03}^{+0.16}$  & 625$\pm50$ & 11 & \cite{ksu06}\\
\enddata
\tablenotetext{a}{A-type stars are defined here as stars with spectral type
B5-A9; earlier B stars are omitted to minimize the possibility of
24\,$\micron$ detection from gaseous disk free-free emission rather than from warm dust in a debris disk.}
\end{deluxetable}

\clearpage
\begin{deluxetable}{cccccc}
\tabletypesize{\scriptsize}
\tablecaption{24\,$\micron$ Excess Frequencies of FGK Stars from {\it
Spitzer}/MIPS Surveys\label{tbl-6}}
\tablewidth{0pt}
\tablehead{
\colhead{} &
\colhead{Excess Frequency} &
\colhead{SpT Range} &
\colhead{Age (Myr)} &
\colhead{Sample Size} &
\colhead{Excess Frequency References}
}
\startdata
Sco Cen\tablenotemark{a}& 0.40$_{-0.08}^{+0.25}$ & G & 16$\pm2$ & 35 & \cite{che05}\\
NGC\,2547 & 0.33$_{-0.08}^{+0.12}$ & F & 30$\pm5 $ & 21 & Gorlova et al. (in prep)\\
{\bf IC\,2391} & {\bf 0.31$_{-0.09}^{+0.13}$} & {\bf F-K4} & {\bf 50$\pm5$} & {\bf 16} & {\bf This paper}\\
Pleiades & 0.09$_{-0.02}^{+0.06}$ & F-K6 & 115$\pm20$ & 53 & \cite{sta05,gor06}\\
Hyades & 0.00+0.03 & G & 625$\pm$50 &51\tablenotemark{b} & \cite{cie06}\\
Field stars & 0.03$\pm$0.02 & F5-K5  & 4000\tablenotemark{c} & 69 & \cite{bry06} 
\enddata
\tablenotetext{a}{Includes only stars from the subgroups Upper Centaurus Lupus and Lower Centaurus Crux.}
\tablenotetext{b}{Results are based on preliminary information from \cite{cie06}, a conference poster. See Appendix for additional comments.}
\tablenotetext{c}{Median age of the sample.}

\end{deluxetable}

\clearpage

\clearpage
\begin{figure}
\includegraphics[angle=+90,width=\columnwidth]{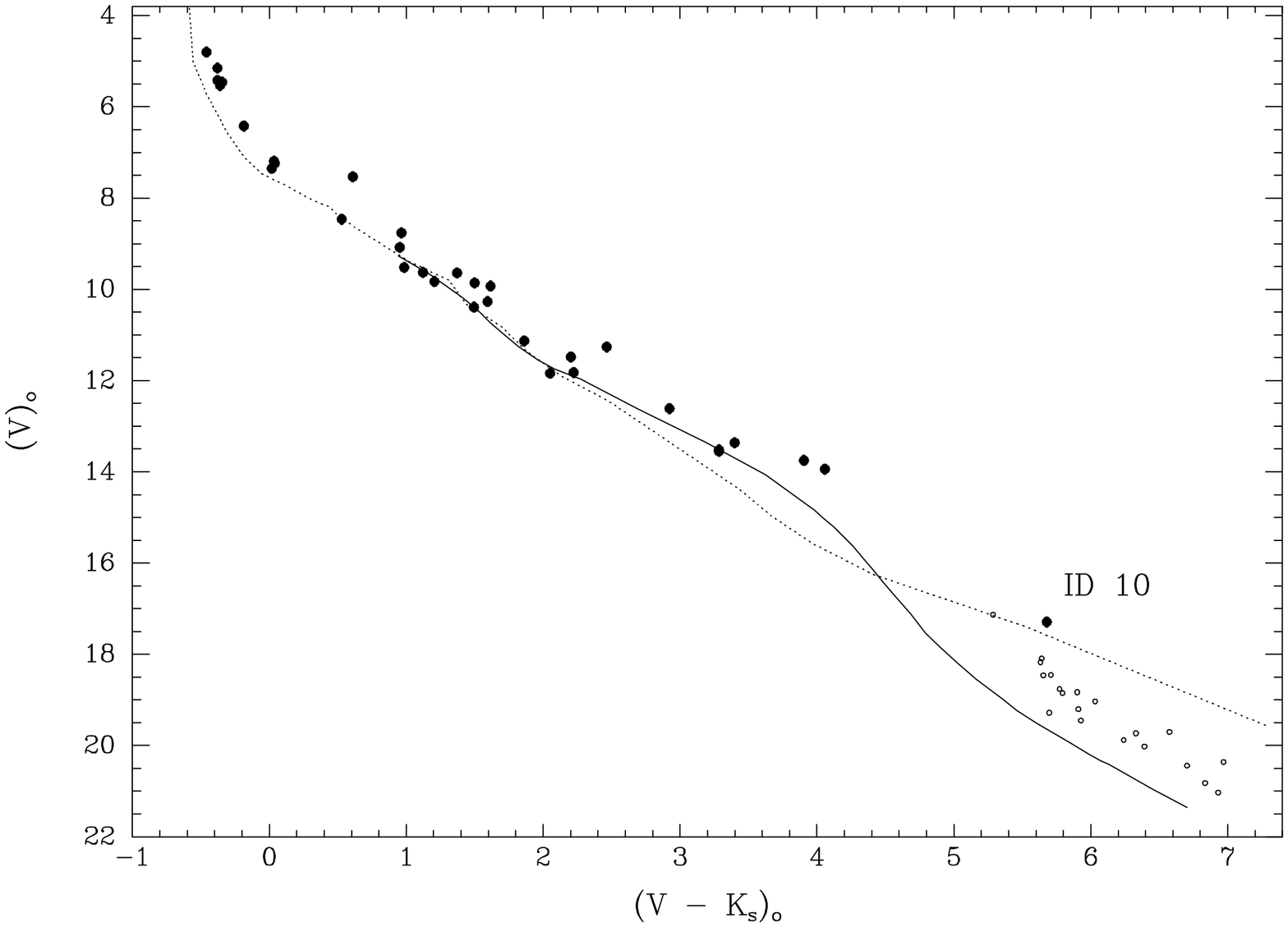}
\caption{Dereddened {\it V} vs {\it V-K$_s$} color-magnitude diagram (CMD) of 33 IC\,2391 cluster members ({\it filled circles}) observed in our 24\,$\micron$ mosaic (Figure \ref{fig1}); stars have been uniformly dereddened using {\it E(B-V)}=0.006 \citep{pat96} and the near-infrared reddening laws of \cite{cam02}. Not included is the brightest star of the cluster $o$\,Velorum (ID\,20) which is saturated at {\it K$_{s}$}.
 Overplotted are 50\,Myr theoretical isochrones from \cite{sie00} ({\it dotted} line) and \cite{bar98} ({\it solid} line) placed at the distance of IC\,2391. Since the models begin diverging at {\it V-K$_{s}$}$\gtrsim$\,4.4, we also plot 22 M4-M7 dwarfs (small {\em open circles}) that are spectroscopically-confirmed cluster members from \cite{byn04} to illustrate the empirical sequence for the coolest known members. The M5 dwarf ID\,10, the faintest member in our sample detected at 24\,$\micron$, appears consistent with membership but as a likely binary.
\label{fig3}} 
\end{figure}

\clearpage
\begin{figure}
   \includegraphics[angle=90,width=\columnwidth]{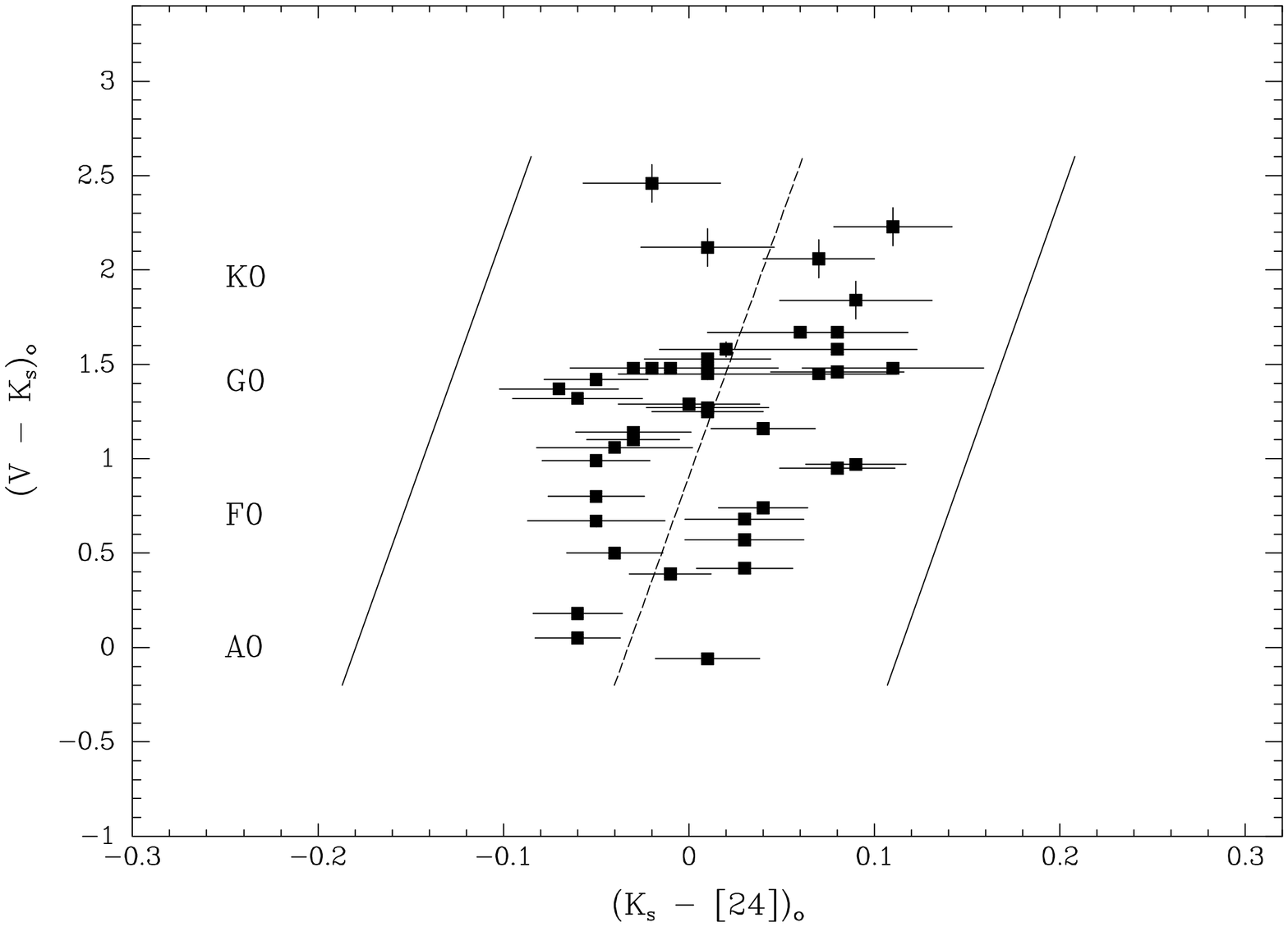}
\caption{Dereddened {\it V-K$_s$} vs {\it K$_s$}-[24] color-color diagram
plotting 57 late-B to mid-K cluster members ({\it solid squares}) of the
Pleiades open cluster possessing no apparent 24\,$\micron$ excess
\citep{sta05,gor06}. The linear fit to the data ({\it central dashed
line}) with a 3$\sigma$ scatter of 0.15\,mag ({\it outer solid lines}) for stars with colors between 0.05\,$\leq$\,{\it V-K$_{s}$}$\leq\,$3.0 is described in \cite{gor06}. We interpret the region within the 3$\sigma$ outer solid lines as the empirical photospheric locus for late-B to mid-K stars.
\label{fig4}}
\end{figure}

\clearpage
\begin{figure}
\includegraphics[angle=+90,width=\columnwidth]{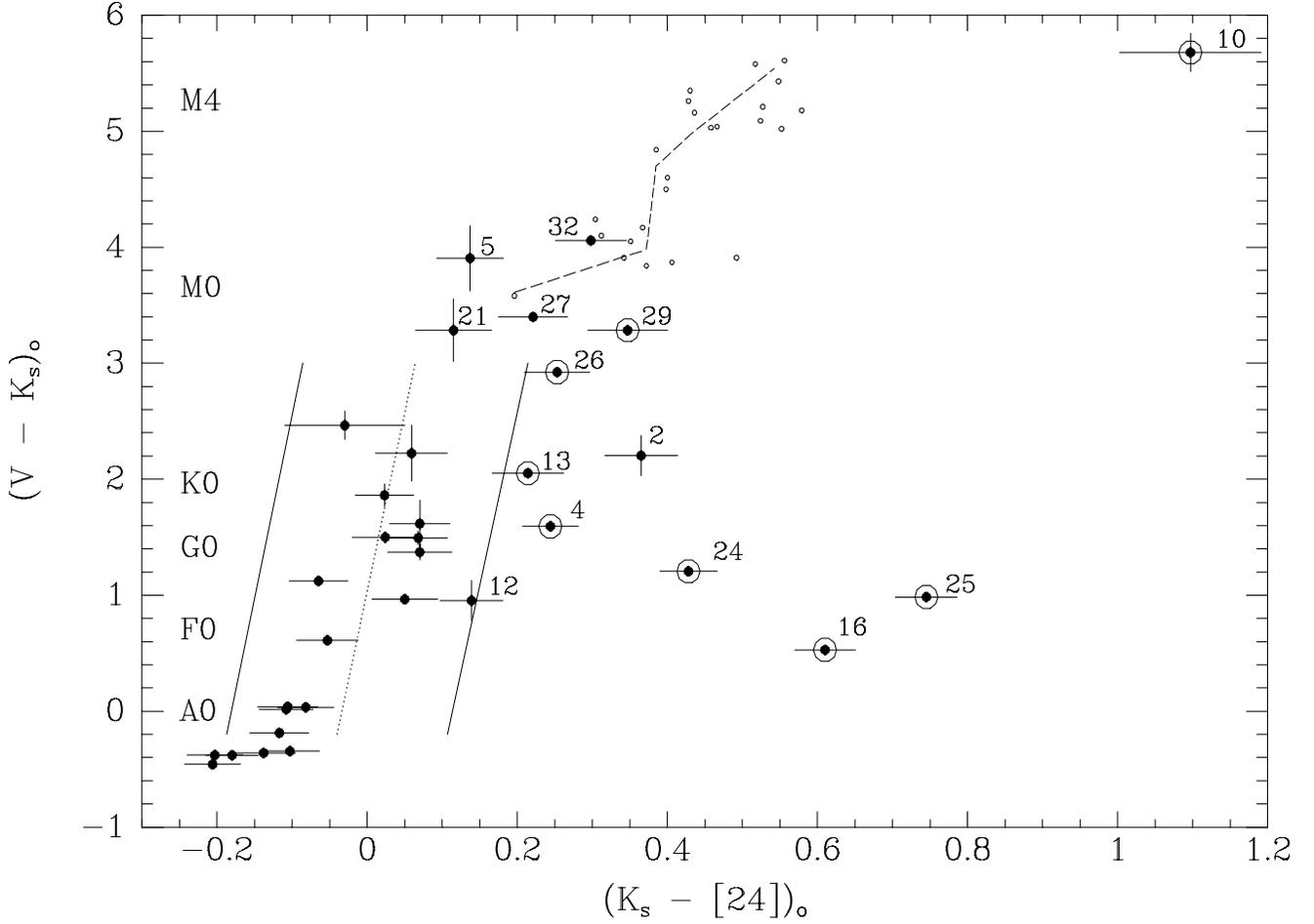}
\caption{Dereddened {\it V-K$_s$} vs {\it K$_s$}-[24] color-color diagram plotting 31 members ({\it filled circles}) of the IC\,2391 open cluster. The brightest star in the cluster, B3IV $o$\,Velorum (ID\,20), is omitted due to saturated K$_s$ photometry. The dereddened Pleiades photospheric locus with its mean ({\it dotted line}) and 3$\sigma$ scatter ({\it solid lines}) from Figure \ref{fig4} is overplotted. Sources redder than the 3$\sigma$ Pleiades relative excess threshold possess {\it K$_s$}-[24] flux ratios in excess of expected photospheric colors and are considered to be debris disk candidates ({\it large open circle}). To estimate the photospheric locus for stars with {\it V-K$_{s}$} colors redder than the Pleiades locus ({\it V-K$_s$}$>$3.0), we plot field M dwarfs ({\it small open circles}) from a 24\,$\micron$ and 70\,$\micron$ investigation by \cite{gau06}. We also plot the theoretical photospheric {\it V-K${_s}$} colors of 50\,Myr stars \citep{sie00} matched to a spectral type/{\it K$_s$}-[24] relation from the Gautier et al. M dwarf sample ({\it dashed line)}. Both profiles, along with a few members from IC\,2391, show a redward turn for photospheres of late-type stars near {\it V-K$_s$}$\sim$\,3.0. ID numbers of some of the cluster members are shown. The mid-infrared excess of ID\,2 may be associated with a background giant (see \S3.4).
\label{fig5}} 
\end{figure}

\clearpage
\begin{figure}
\includegraphics[angle=+90,width=\columnwidth]{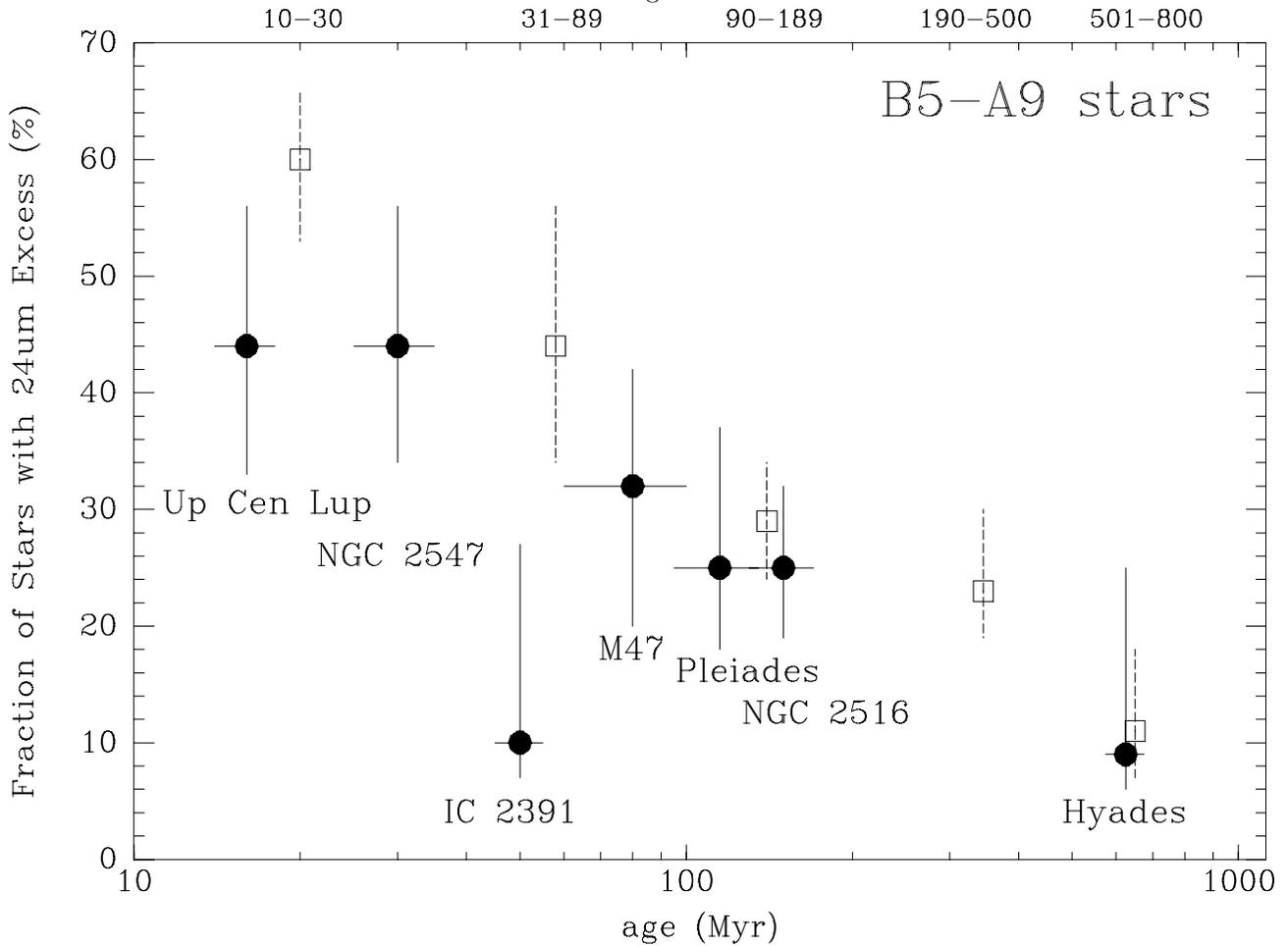}
\caption{Fraction of A-type stars (B5-A9) with 24\,$\micron$ excess as a
   function of age. Plotted as {\it filled circles} are the excess frequencies from {\it Spitzer}/MIPS 
   observed open clusters and a stellar association (data and references listed in Table \ref{tbl-4}). Also shown in {\it open boxes} are mean 24\,$\micron$ excess frequencies from age bins (listed across the top of the Figure) from the combined MIPS-only surveys of \cite{rie05} and \cite{ksu06}. Error bars are 1$\sigma$ binomial distribution uncertainties and age uncertainties are taken from the cluster references in Table \ref{tbl-4}. In all cases a 15$\%$ relative excess threshold was used. Note that the IC\,2391 excess frequency appears comparatively low for its age. 
\label{fig7}} 
\end{figure}

\clearpage
\begin{figure}
\includegraphics[angle=+90,width=\columnwidth]{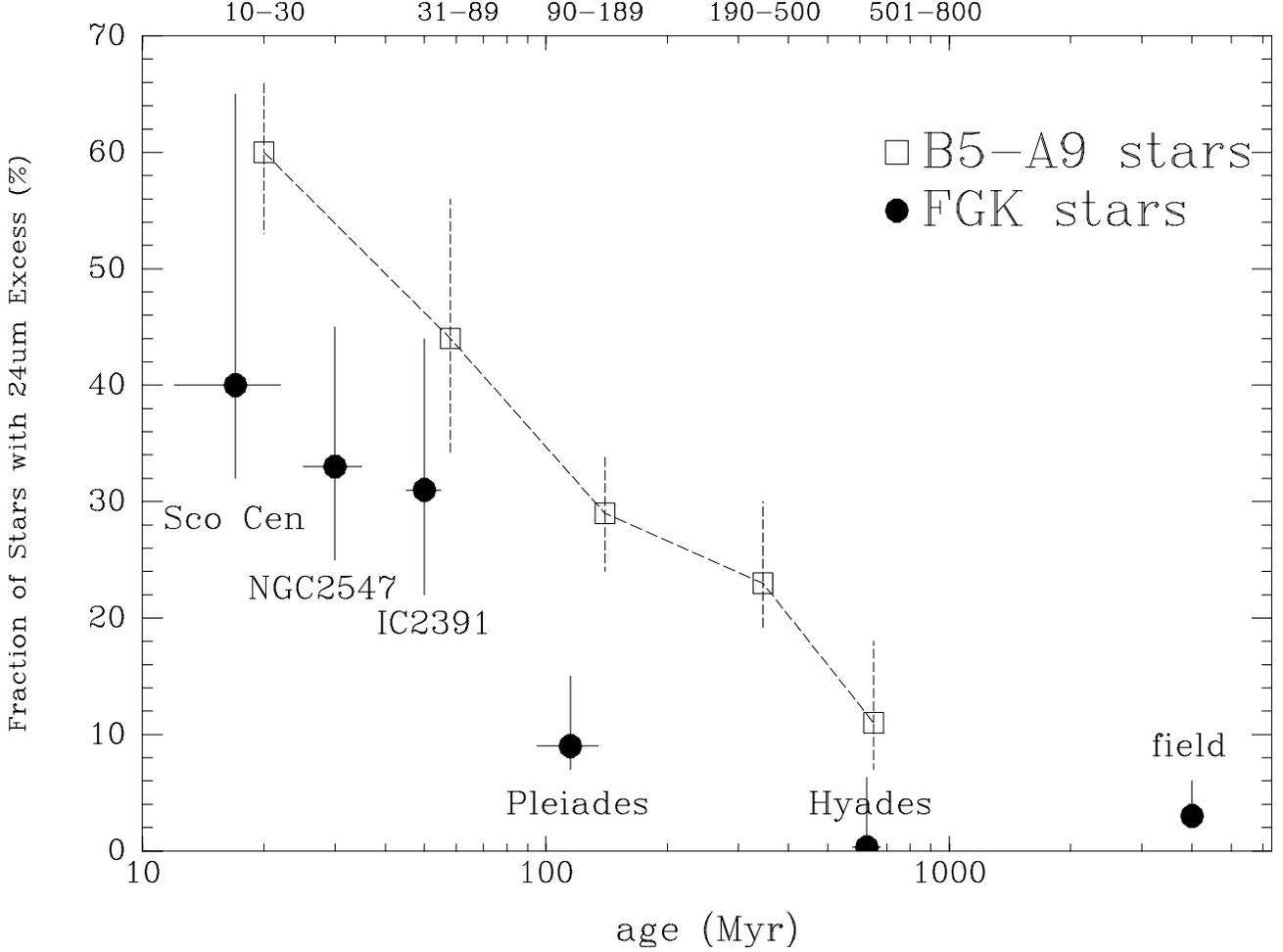}
\caption{Fraction of stars with spectral types F, G, and/or K with 24\,$\micron$ excess as a
   function of age. Plotted as {\it filled circles} are the excess frequencies of these stars from {\it Spitzer}/MIPS 
   observed clusters and an association (references given in Table
   \ref{tbl-6}; additional comments are made in Appendix B). As a
   comparison, we also plot the mean 24\,$\micron$ excess frequencies of
   A-type stars ({\it open boxes}) from the combined MIPS-only data of
   \cite{rie05} and \cite{ksu06} as shown in Figure \ref{fig7} (we connect the A-type points with a {\it dashed line} to help distinguish the two populations). In all cases, a 15$\%$ relative excess threshold was used. Vertical error bars
   are 1-sigma binomial distribution uncertainties and age
   uncertainties are taken from the cluster references listed in Table
   \ref{tbl-6}. Despite probing different annular regions, the decline in the excess frequency of the FGK stars suggests a decline in the collision rate between planetesimals with
   stellar age similar to that of the more massive A-type stars. 
\label{fig8}} 
\end{figure}

\clearpage
\begin{figure}
\includegraphics[angle=+90,width=\columnwidth]{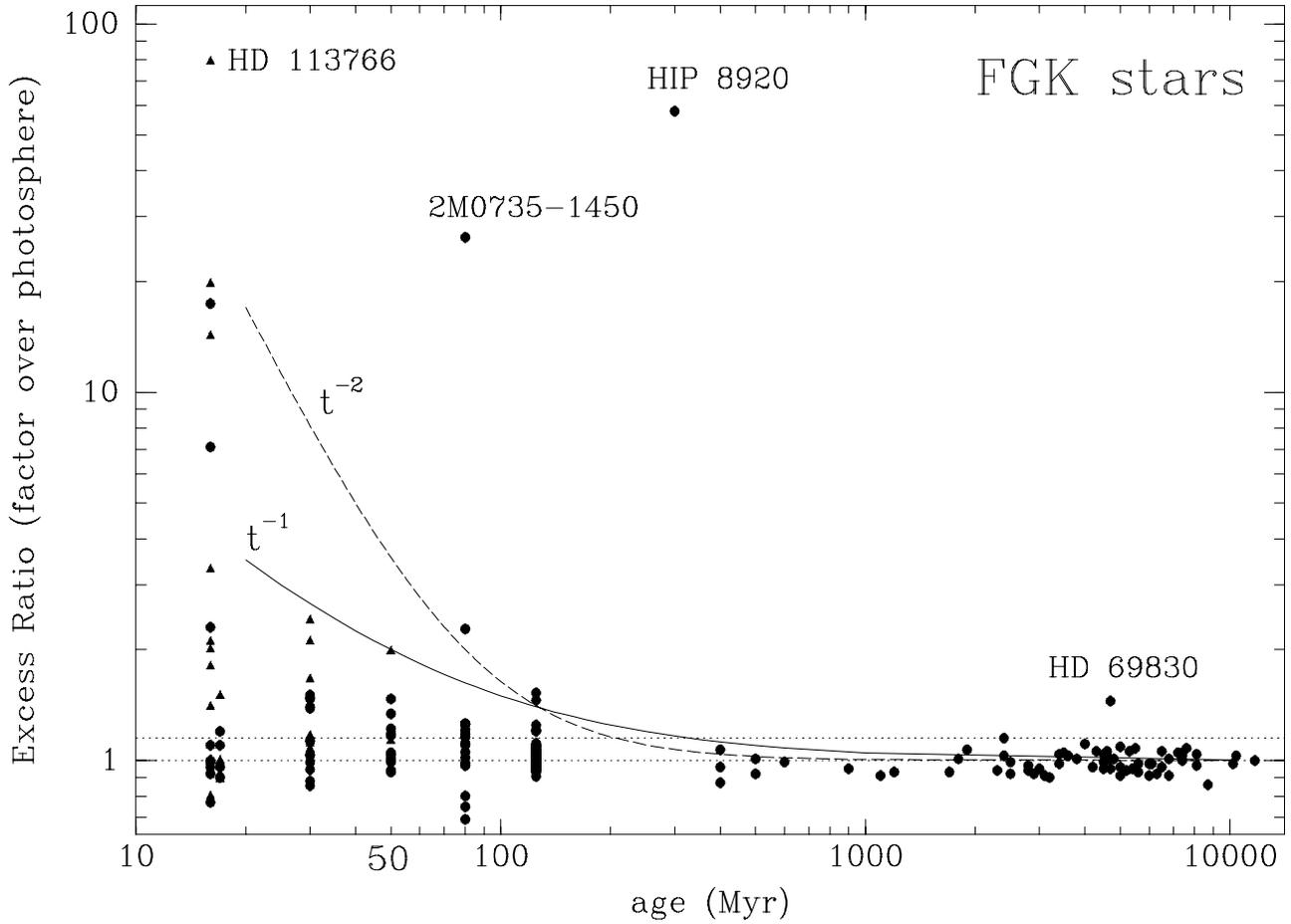}
\caption{24\,$\micron$ excess ratio vs age for FGK stars. The excess
   ratio is the measured flux density to that expected from the
   stellar photosphere alone; a value of 1 represents no excess ({\it
   lower horizontal dotted line}). The {\it upper horizontal dotted
   line} represents the 1.15 relative excess threshold used in this
   study. F0-F4 stars are shown as {\it filled triangles} while F5-K7
   stars (``solar-like'') appear as {\it filled circles}. The {\it
   solid curve} is an inverse time dependence and the {\it dashed
   curve} is inverse time-squared. All of
   the points have been observed with {\it Spitzer} at 24\,$\micron$
   and, with a few noted exceptions, are part of the disk investigations listed in Table
   \ref{tbl-6}. Additions are M47 data \citep{gor04}, the 30\,Myr
   HD\,12039 \citep{hin06}, and the 300\,Myr HIP\,8920
   \citep{son05}. Omitted from this figure is HD\,152404 from Upper
   Centaurus Lupus \cite[$\sim$\,17\,Myr, F5V;][]{che05} with a
   reported excess ratio of 202. The size of the excess coupled with
   its age suggests the possibility that the large excess may be due
   to a long-lived primordial disk. IC\,2391's data are shown above its
   labelled age of 50\,Myr. 
\label{fig12}} 
\end{figure}

\clearpage
\appendix
\section{Literature Sources Inconsistent with Cluster Membership}
The following sources are classified in the literature as possible or probable cluster members, but we conclude from this investigation that their properties are inconsistent with membership.

\noindent {\bf VXR PSPC 31}\\
\noindent Source at 08\,41\,11.0 -52\,31\,46.0 is 0.47\,mag below IC\,2391 single-star sequence in Figure \ref{fig3}. In addition, both the Tycho-2 and UCAC2 proper motions exclude it as a member with high significance ($\chi^{2}/\nu\simeq$\,42/2). In addition, colors and spectral type indicate evidence of reddening which is inconsistent with the overall cluster reddening. Despite evidence of youth \citep{ran01}, the source is likely to be a young background object rather than a cluster member.

\noindent {\bf HD\,74517}\\
\noindent Tycho-2 proper motion is well-constrained but largely inconsistent with the \cite{rob99} cluster mean ($\chi^{2}/\nu\simeq$\,320/2). The nearly zero proper motion suggests that the source is likely to be an interloping A star rather than a cluster member. 

\noindent {\bf HD\,74665}\\
\noindent The proper motion is known to high accuracy and the Hipparcos, Tycho-2, and UCAC2 proper motions exclude the star as a kinematic member:
                  ($\chi^{2}/\nu\simeq$\,115/2, 81/2, 107/2, respectively). Source is likely an interloping A star rather than a cluster member.

\section{Comments on Individual Clusters}

\noindent {\bf Scorpius-Centaurus OB association (16-17\,Myr, 132$\pm$14\,pc)}\\
\noindent \cite{che05} have described a survey for excesses in this association's two oldest
subgroups, Upper Centaurus Lupus ($\sim$17\,Myr) and Lower Centaurus Crux
($\sim$16\,Myr). Initial results show a 24\,$\micron$ excess frequency of $\sim$40$\%$
(14/35 using a relative excess threshold $\geq$1.15, private communication Christine Chen). Chen et al. state that the frequency could potentially be 40$\%$ higher if presumed interlopers are identified and removed from their proper motion selected sample. The large reported upper error bar in Table \ref{tbl-6} is due to basing the uncertainty on this possible contamination. Age estimates are from \cite{mam04}; distance is reported as typical stellar distances from the 3 subgroups \citep{che05}.

Regarding the A-type stars in Upper Centaurus Lupus observed at 24\,$\micron$ with {\em Spitzer} by \cite{ksu06}, we used a 15\% threshold in determining the number of excess sources so as to be consistent in our treatment of all the surveys. \cite{ksu06}, however, with improved photometry and Kurucz photospheric model fitting have reduced the threshold to 6\% for the {\em Spitzer} A-type stars in their sample. Consequently, they measure an excess frequency of 56\% (9/16) rather than the 44\% (7/16) we report in Table \ref{tbl-4} and shown in Figure \ref{fig7}.

\noindent {\bf NGC\,2547 (30$\pm$5\,Myr, 450$\pm$45\,pc)}\\
\noindent Using the Pleiades photospheric locus as the relative excess threshold and a larger list of cluster members, Gorlova et al. (in prep) have improved upon the number of sources with 24\,$\micron$ detections from \cite{you04} to now include F stars.

\noindent {\bf M47 (80$\pm$20\,Myr, 450$\pm$50\,pc)}\\
\noindent The scatter of the {\it K$_{s}$}-[24] color for FGK stars is relatively large with some sources appearing blue-ward of the Pleiades photospheric locus. While there were both F and G stars detected in the 24\,$\micron$ investigation of M47 \citep{gor04}, the photometry was obtained during the {\it Spitzer} early checkout period during telescope commissioning and data analysis techniques were still being optimized. Consequently, identifying excess sources among the FGK stars with a 15$\%$ threshold cannot yet be done with great confidence and hence we do not use the cluster in our evolution analysis of FGK disk frequencies. 

While we do not utilize the photometry of the solar-like stars in
determining a debris disk frequency, we include M47 data in the excess
ratio evolution (Figure \ref{fig12}) since we are interested in the range of excesses rather than the frequency.

\noindent {\bf Hyades (625$\pm$50\,Myr, 46.3$\pm$0.3\,pc)}\\
\noindent 
We include unpublished preliminary MIPS/24\,$\micron$ results of the 625\,Myr Hyades open cluster \cite[conference poster;][]{cie06} who report {\em no} sources with excess ratios clearly above $\sim$\,25$\%$. At the 15$\%$ level, there is evidence for one borderline excess source from a re-reduction of the public Hyades 24\,$\micron$ data during this investigation. Age estimate is from \cite{per98}.

\noindent {\bf Pleiades (115$\pm$20\,Myr, 135$\pm$3\,pc)}\\
\noindent 
Fifty-three members of the Pleiades with spectral types between B8 to K6 have been analyzed by \cite{sta05} and \cite{gor06} identifying five with evidence of debris disks. References for cluster age and distance are taken from those within \cite{gor06}.

\noindent {\bf Field stars}\\
\noindent 
Targeting 69 older, nearby field solar-type stars with median age $\sim$4\,Gyr, \cite{bry06} only found 2 objects with 24\,$\micron$ excess $\geq$\,15$\%$ above the photosphere.

\end{document}